\documentclass[preprint,12pt,authoryear]{elsarticle}

\usepackage{amssymb}
\usepackage{amsmath}

\usepackage{lineno}

\usepackage{verbatim}

\journal{Icarus}

\begin{document}

\begin{frontmatter}



\title{Alkali recondensation into chondrules} 


\author[MNHN]{Emmanuel Jacquet} 
\author[CRPG]{Yves Marrocchi}
\author[IPGP]{S\'{e}bastien Charnoz}


\affiliation[MNHN]{organization={Institut de Min\'{e}ralogie, de Physique des Mat\'{e}riaux et de Cosmochimie, Mus\'{e}um national d'Histoire naturelle, CNRS},
            addressline={CP52, 57 rue Cuvier}, 
            city={Paris},
            postcode={75005}, 
            country={France}}

\affiliation[CRPG]{organization={Centre de Recherches P\'{e}trographiques et G\'{e}ochimiques},
            addressline={}, 
            city={Vandoeuvre-l\'{e}s-Nancy},
            postcode={54501}, 
            country={France}}

\affiliation[IPGP]{organization={Institut de Physique du Globe de Paris, Universit\'{e} de Paris, CNRS},
            addressline={1 rue Jussieu}, 
            city={Paris},
            postcode={75005}, 
            country={France}}

\begin{abstract}
While sub-mm melt droplets should rapidly lose alkali elements in a vacuum at liquidus temperatures, chondrules are only modestly depleted in them (by less than one order of magnitude). The detection of sodium in olivine cores has previously suggested very high saturating partial pressures of gaseous sodium, but we show that alkalis were lost during heating and recondensed at lower temperatures, essentially in the present-day chondrule mesostases. This recondensation was accompanied by mass-dependent enrichment in light isotopes (for multi-isotope alkalis such as K and Rb), but its limited extent indicates a cooling acceleration (or "quenching"). The isotopic fractionation also constrains the ratio of the chondrule density and the cooling rate prior to the quench around $10^{-6}\:\mathrm{kg.m^{-3}.K^{-1}.h}$ suggesting densities above $\sim 10^{-6}\:\mathrm{kg/m^3}$.  In a nebular context, this is achievable by radial and vertical concentrations near pressure bumps.
\end{abstract}




\end{frontmatter}



\section{Introduction}
\label{intro}

  Chondrules are sub-millimeter spheroidal droplets ubiquitous in primitive meteorites (or chondrites). It is largely accepted that chondrules result from magmatic processes involving maximum temperatures in the range 1600-2000 K \citep{Jonesetal2018}, but their formation conditions and the physical process(es) operating are still hotly debated (see \citet{Marrocchietal2024ISSI} for a review). Mounting evidence of calcium-aluminum-rich inclusions (CAI) or amoeboid olivine aggregates (AOA) among their precursors, either in the form of identifiable relicts \citep[e.g.][]{Krotetal2006,Marrocchietal2018,Marrocchietal2019relicts,Pirallaetal2021} or bulk chemical \citep[e.g.][]{MisawaNakamura1988,JacquetMarrocchi2017} or isotopic \citep{Gerberetal2017,Schneideretal2020} signatures, suggests "nebular scenarios" where pristine nebular solid aggregates were melted, e.g. by shock waves \citep{Deschetal2005,Boleyetal2013}, lightning discharges \citep{DeschCuzzi2000,JohansenOkuzumi2018}, or intermittently exposed magma oceans \citep{HerbstGreenwood2019}.

  Yet, chondrules also carry signatures of high solid densities in their formation regions, which have given some support to "planetary scenarios" involving impacts \citep{Asphaugetal2011,SandersScott2012}. The frequency of compound chondrules \citep{GoodingKeil1981,SekiyaNakamura1996,Jacquet2021}, the Fa contents of chondrule olivine \citep{Grossmanetal2012,Tenneretal2015}, and the limited Mg isotopic fractionation of bulk chondrules \citep{CuzziAlexander2006} indicate dust densities above $\sim 10^{-6}$kg/m$^3$ with a dust/gas ratio enhanced by 1-3 orders of magnitude above solar \citep{Jacquetetal2024}. Conceivably though, a combination of radial concentration, e.g. around pressure maxima, commonly invoked to explain rings in present-day protoplanetary disks, and settling in a nonturbulent nebula could bring dust concentrations to such levels near the disk midplane, hovering around the threshold for streaming instability and planetesimal accretion \citep{Jacquetetal2024}.

  However, a further constraint threatens to make this hard-fought solution hopelessly inadequate: the abundance of alkali elements, in particular sodium (the most abundant one). Indeed, these moderately volatile elements would be expected to be rapidly lost (within a minute) in vacuum upon heating \citep{Tsuchiyamaetal1981,Yuetal2003,FedkinGrossman2013}. They certainly are depleted in type I chondrules (that is, reduced chondrules, with silicate Fe\#$\equiv \rm Fe/(Fe+Mg)<$10 mol\%), but by less than one order of magnitude, and type II (oxidized) chondrules show almost chondritic levels \citep{Hewins1991}. While very brief heating (with unimpeded radiative cooling) could avert sodium loss \citep{Hewins1991,RubinWasson2005, Baeckeretal2017}, estimated cooling times are generally longer by a few orders of magnitude \citep{Jonesetal2018}, and the multiplicity of the purported flash heating events \citep{Baeckeretal2017} would further thwart their individual advantage in sodium retention \citep{Jacquetetal2015LL}. Still, recondensation upon cooling could make up for the initial loss \citep[e.g.][]{Alexanderetal2000} but the detection of sodium in chondrule olivine cores has suggested that chondrules behaved as essentially closed systems for alkalis below \textit{liquidus} temperatures \citep{Alexanderetal2008}.  Barring very high (atmospheric level!) total pressures which could suppress the diffusion of evaporated alkalis away from the chondrules, this would require dust densities in excess of $10^{-3}$kg/m$^3$, six orders of magnitude above conventional "minimum mass solar nebula" estimates, to allow saturation of the ambient gas in gaseous sodium with a subordinate evaporation fraction per chondrule.  This would place chondrule formation in self-gravitating regions, or in definitely "planetary" environments \citep[e.g.][]{Dullemondetal2016}.

  While early chondrule-scale analyses had found no resolvable K isotopic fractionation \citep{Alexanderetal2000,AlexanderGrossman2005}, chondrite bulk isotopic measurements have revealed \textit{light} isotope enrichments in K and Rb correlated with the abundance of chondrules (for carbonaceous chondrites), suggestive of recondensation rather than evaporation \citep{Nieetal2021, Koefoedetal2023}.  Recondensation has appeared also more graphically suggested by alkali-zoned chondrules, with alkalis concentrated near their margins \citep[e.g.][]{Matsunamietal1993,Liboureletal2003,Nagaharaetal2008}, even though crystal-melt partitioning or parent body processes were also envisioned \citep{Grossmanetal2002}. The time is thus ripe to revisit the alkali 
 retention problem in a synthetic manner (glass composition, zoning in olivine and mesostases, isotopic constraints), leveraging the data accumulated in the various chondrite groups in the literature. We show that chondrules were definitely open systems for alkalis and revise accordingly the density constraints brought by them.

  We will first qualitatively review the cosmochemical evidence for alkali recondensation (from bulk composition, olivine zoning and glass inclusions) in section \ref{evidence}. We will then expose the thermodynamics of this recondensation in section \ref{thermodynamics}, supporting some statements made in advance in the previous section. Yet this, by itself, will not allow conclusions on chondrule-forming conditions, absent a determination of the closure temperature. Section \ref{kinetics} will thus deal with the kinetics of alkali recondensation, in particular the isotopic effects (for multi-isotope alkalis). Only then will section \ref{context} draw and discuss constraints on the astrophysical setting of chondrule formation.



\section{Evidence for recondensation}
\label{evidence}

\subsection{Data sources}
Mesostasis and bulk chondrule data were compiled from the literature for: ordinary \citep[chiefly LL3.0 to LL3.2;][]{DeHart1989,Jones1990,Jones1994,Jones1996,Huangetal1996,Grossmanetal2002,Liboureletal2003,Tachibanaetal2003,Nagaharaetal2008,Alexanderetal2008,Kitaetal2010,Villeneuve2010,Jacquetetal2015LL,Berlin2009,Hewinsetal2012,Grossmanetal2002}, EH \citep{Grossmanetal1985,Ikeda1989,Schneideretal2002,Jacquetetal2015EC,Pianietal2016,DouglasSongetal2025}
,  CR \citep{Krotetal2004,Mathieu2009,Berlin2009,WassonRubin2010,IchikawaIkeda1995,Jacquetetal2012CC,SmithJones2024}, CV \citep[3.0 to 3.1 in the \citet{JacquetDoisneau2024} subtypes;][]{
Kracheretal1985,KimuraIkeda1997,KimuraIkeda1998,Krotetal1998,Jacquetetal2012CC},  CO \citep[3.0 to 3.2;][]{RubinWasson1988,Jones1992,Berlin2009,Tenneretal2013,Friendetal2016,Jacquet2021}, CM \citep{Ikeda1983,Olsen1983,Kimuraetal2020}, CH \citep{Nakashimaetal2024}, C3-an Acfer 094 \citep{Ushikuboetal2012} and Dar al Gani 055 \citep{VarelaKurat2009}, C2-an Essebi \citep{VarelaKurat2009}, CB$_b$ \citep{Krotetal2010}, CH/CB$_b$ \citep{Krotetal2007}. The CM, CO and Acfer 094 data will be grouped in the "CMO" clan \citep{Jacquet2022}, while "CHB" will refer to CH, CB and Isheyevo (which likely derive from a single parent body; e.g. \citet{Mahleetal2024}).

\subsection{Bulk chondrule compositions}

\begin{figure}
\centering
\includegraphics[width=\textwidth]{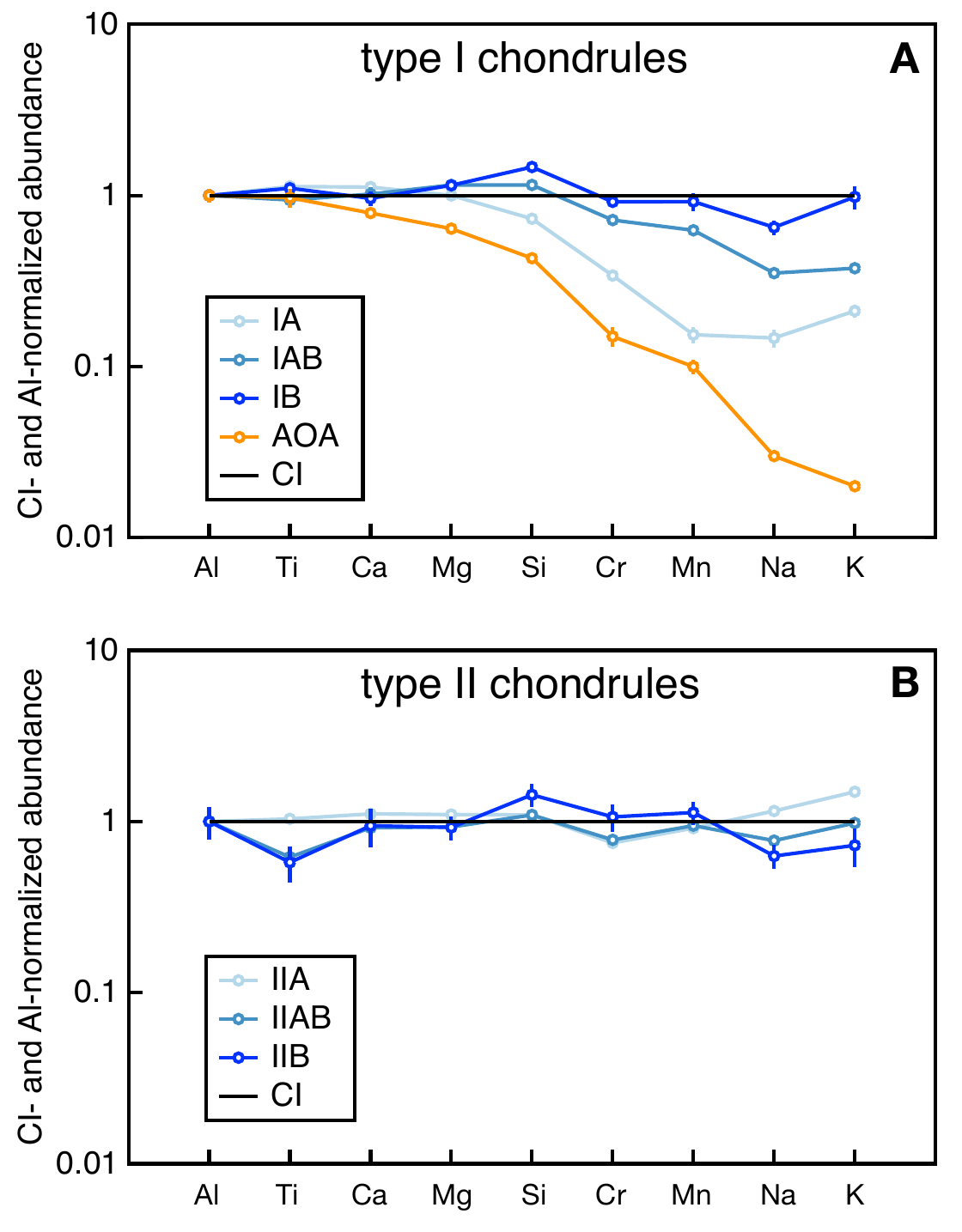}
\caption{Abundance patterns for elements arranged in order of increasing volatility 
 for different types of chondrules in ordinary chondrites (type I and II in panels A and B respectively). The numbers of averaged chondrules were 19 (IA), 17 (IAB), 9 (IB), 7 (IIB), 45 (IIAB), 41 (IIA). Error bars are one standard error of the mean. The profile for amoeboid olivine aggregates \citep[AOA;][]{Ruzickaetal2012} is also shown.}\label{pattern}
\end{figure}

\begin{figure}
\centering
\includegraphics[width=\textwidth]{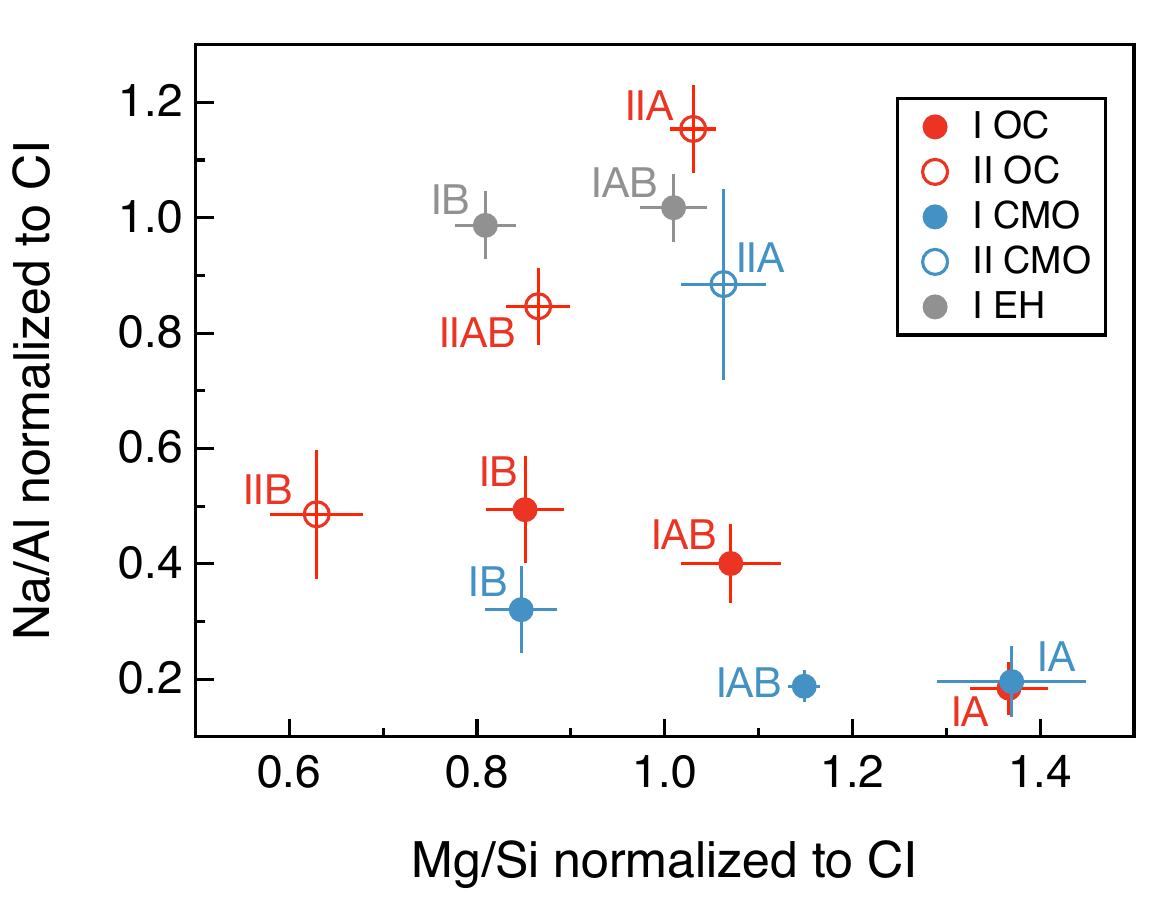}
\caption{Mean Na/Al versus Mg/Si ratios for different chondrule types in ordinary, enstatite and CMO chondrites. Error bars are one standard error of the mean. The numbers of averaged chondrules were the same as Fig. \ref{pattern} for OCs, and, for CMO, 70 (IAB), 18 (IB), 8 (IIA), and for EH, 17 (IAB) and 19 (IB).}\label{Na_vs_Mg}
\end{figure}

\begin{figure}
\centering
\includegraphics[width=\textwidth]{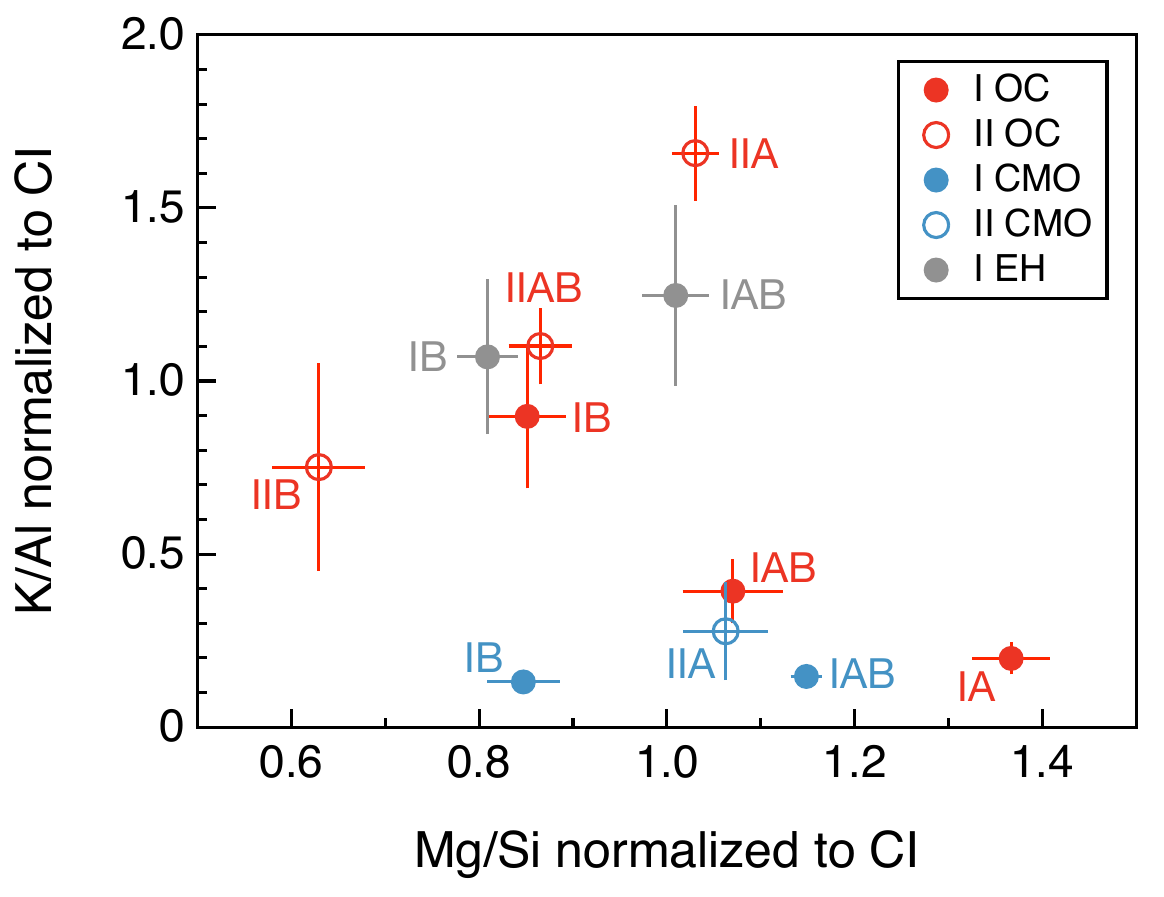}
\caption{Mean K/Al versus Mg/Si ratios for different chondrule types in ordinary, enstatite and CMO chondrites. Error bars are one standard error of the mean. Same numbers of averaged chondrules as Fig. \ref{Na_vs_Mg}.}\label{K_vs_Mg}
\end{figure}

  Major and minor lithophile element patterns for bulk chondrules are shown in Fig. \ref{pattern}, for ordinary chondrites, and alkalis are illustrated for different chondrite groups in Fig. \ref{Na_vs_Mg} and \ref{K_vs_Mg}. Averages are by types, with an "A" suffix indicating olivine dominating over pyroxene by at least 9:1 (in volume), "B" the contrary relationship, and "AB" the intermediate mineralogies. These are arbitrary divisions in a continuum. There is however a true dichotomy between type I and type II, even though, for ordinary chondrites, IB and IIB chondrules seem to make the connection, and the exact conventional divide in Fe\# may vary among authors, as reviewed by \citet{Rubin2024}. 
In carbonaceous chondrites, type II chondrules are essentially restricted to type IIA (there is virtually no Fs content above 10 mol\% in the histograms of \citet{Wood1967}), possibly related to their more oxidizing conditions (which promote olivine at the expense of pyroxene; \citet{JacquetDoisneau2024}), so the hiatus is clear-cut. In enstatite chondrites, there are none. 
  
  The type I chondrules are increasingly volatile- (and alkali-)depleted in the sequence IB-IAB-IA. This is strong evidence for an open system behavior, since pyroxene itself seem to result from interaction with the gas in type I chondrules. Indeed pyroxene is usually concentrated at their margins \citep{Tissandieretal2002,Liboureletal2006,Baroschetal2019} and mesostasis compositions (with anticorrelated SiO$_2$ and Al$_2$O$_3$) reflect dilution by recondensing SiO rather than closed-system subtraction of olivine \citep{Liboureletal2006}. The silica enrichment of the melt, which did not readily diffuse inward, promoted the formation of pyroxene at the expense of (often poikilitically enclosed) olivine \citep{Liboureletal2006,Soulieetal2017}. So alkalis in particular likely recondensed along with Si \citep{Liboureletal2003}.  \citet{Jacquet2021} showed that volatile elements in (type I) compound chondrule components correlated better (Na/Al has a correlation coefficient of 0.97) 
 than refractory elements, which reflected precursor nugget effects, while the former were buffered by the gaseous environment. The CI-normalized Mn/Na ratio increases in the sequence IA-IAB-IB, presumably because of decreasing equilibration temperature \citep{Sossietal2025}, but is lower in EH chondrules than in OC chondrules ($0.3\pm 0.1$ vs. $1.2\pm 0.2$, one standard error of the mean, n=33 and 46 respectively), reflecting lower oxygen fugacity in the former \citep{Sossietal2025}.

  For type II chondrules, their near-chondritic bulk compositions provide less evidence in themselves for open system, even though the IIB and IIAB are somewhat volatile-depleted (possibly owing to less oxidizing conditions, see \citet{Jacquetetal2015LL} and the next sections). At any rate, the Na-Al correlation previously reported for reconstructed 2D bulks should no longer be interpreted in terms of an albite or oligoclase precursor \citep[in a closed-system scenario, also ;][]{Hewins1991,Jacquetetal2015LL}. Mesostasis itself, which is the main carrier of Na and Al, shows no such correlation \citep[e.g.][]{Hewinsetal2012}, and that observed for reconstructed bulks is then likely an artifact of varying proportions of mesostases appearing in the 2D sections analyzed \citep{HezelKiesswetter2010}. It is just a matter of coincidence that the CI chondritic Na/Al ratio (0.68; \citet{Lodders2003}) is close to that of albite--indeed, feldspar in equilibrated ordinary chondrites, which is the main carrier of both Na and Al, has an oligoclase composition (e.g. Ab$_{86}$An$_{10}$Or$_4$ for equilibrated LL chondrites, hence a Na/Al ratio of 0.78; \citep{BrearleyJones1998}). 
Interestingly, most type II (in particular IIA) chondrules in ordinary chondrites show suprachondritic K/Al ratios (with an average CI-normalized K/Al of 1.3$\pm$0.1, one standard error of the mean, n=96; Fig. \ref{K_vs_Mg}, \ref{K_vs_SiO2}), which cannot be accounted for by incomplete recondensation in a closed chondritic reservoir. Unless K-depleted type II chondrules were underrepresented in our compilation, this suggests an influx of K, e.g. lost from type I chondrules if formed nearby \citep[e.g.][]{Ruzicka2012}.

\subsection{Olivine zoning}

\begin{figure}
\centering
\includegraphics[width=\textwidth]{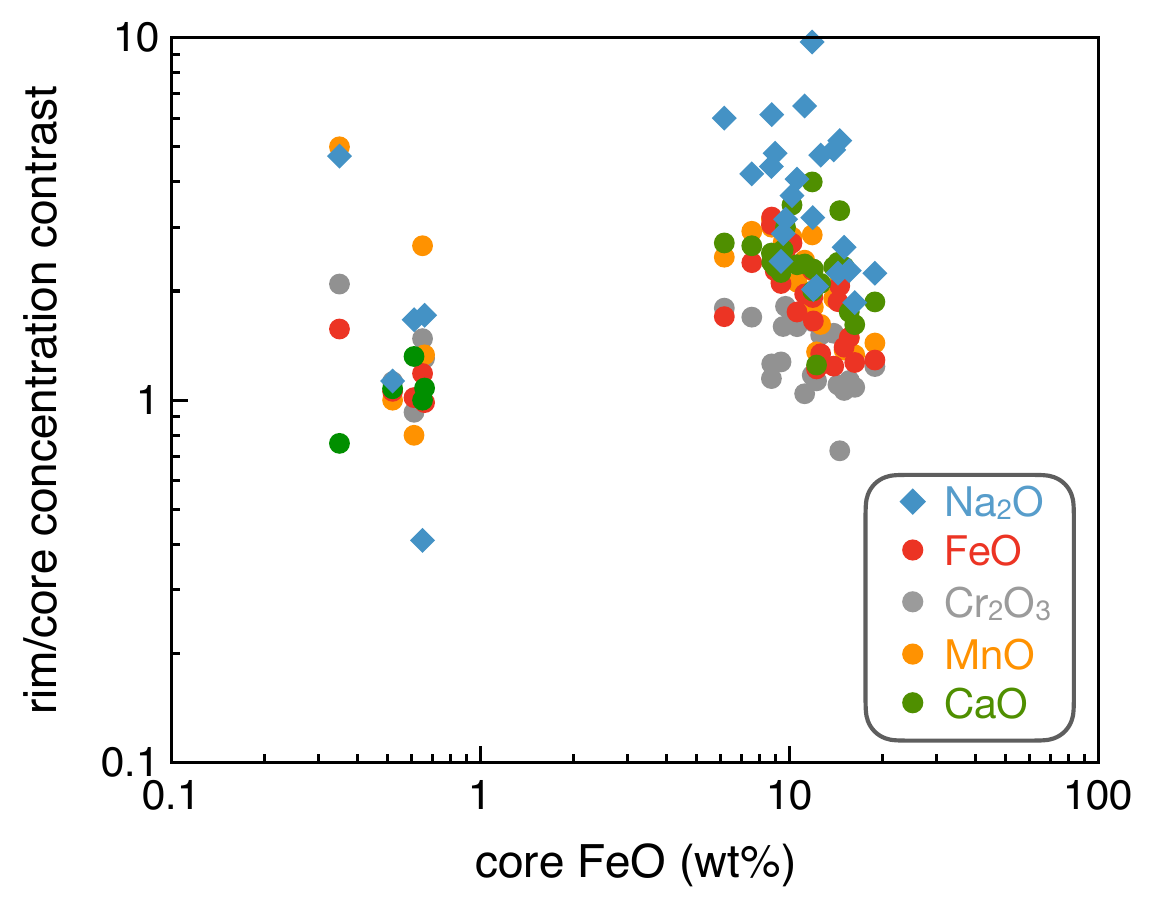}
\caption{Minor element concentration ratio between rim and core olivine in ordinary chondrites \citep{Alexanderetal2008,Hewinsetal2012}, as a function of core FeO content.}\label{zoning_Ol}
\end{figure}

  \citet{Alexanderetal2008} claimed a closed-system behavior for Na below liquidus temperatures because they detected Na even in the core of olivine crystals. Forsteritic relicts grains \citep[e.g.][]{Pintoetal2024} concealed by Fe-Mg interdiffusion in back-scattered electron images,  as speculated by \citet{Jacquetetal2024}, would not be a solution, since those visible are distinctly Na-depleted compared to surrounding host olivine \citep{Alexanderetal2007}
. \citet{Alexanderetal2008} could fit the zoning profiles with a fractional crystallization model (i.e. with negligible diffusion). This, in fact, holds for type II chondrules rather than type I chondrules which show little zoning (Fig. \ref{zoning_Ol})--and for which a closed system would imply \textit{higher} dust densities, despite their lower recorded oxygen fugacity, because of their higher liquidus temperatures \citep{Alexanderetal2008}. 

  However, this approach neglected the positive correlation of the olivine/melt Na partition coefficient for Na with FeO 
\citep{Mathieu2009}\footnote{We may neglect contrary claims by \citet{Jacquetetal2015LL} based on the activity coefficient of Na in the melt, for this is obviously not the only factor in the partition coefficient when writing the action mass law for the substitution reaction, and should thus not supersede the empirical evidence brought by \citet{Mathieu2009}.}. If one sticks to the fractional crystallization hypothesis, one must allow for a post-liquidus loss of half of the Na from type II chondrules, followed, for type IIAB, by more or less complete recovery \citep{Hewinsetal2012}. This could evoke impact plume expansion scenarios, where cooling is counteracted by the density decrease for Na retention \citep{Dullemondetal2016}. Yet, as mentioned in the previous subsection, type IIA chondrules seem to have lost overall \textit{less} Na (with respect to a chondritic precursor, when normalized to refractory Al), on average, than type IIAB chondrules.

  In fact, crystallizing olivine may not have been closed to diffusion. Diffusivities for Na are of the same order of magnitude than for other minor elements (Cr, Mn, Ca, Fe-Mg) in the experiments of \citet{SpandlerONeill2009}, around 10$^{-15}$ m$^2$/s 
  at 1300 °C, even though profiles through forsteritic relict grains suggest it is somewhat slower than the others \citep{Alexanderetal2007}.
 For reference, diffusivities of 10$^{-15}$ m$^2$/s would allow core values within one order of magnitude of the rim within a few hours for grains of a few tens of $\mu$m \citep{Crank1975}
.  Since \citet{Miyamotoetal2009} did find varying degrees of diffusional modifications of Fe-Mg zoning profiles in Semarkona type II chondrules
, Na should have significantly diffused as well. Zoning is stronger for Na than for other elements (Fig. \ref{zoning_Ol}), consistent with more sluggish diffusion and/or more contrasted original profile (e.g. Na-free original cores).  Zoning is nevertheless less pronounced for the most fayalitic olivines, perhaps as a result of increased diffusivity for higher Fa contents and/or oxygen fugacities (as seen for other elements, e.g. \citet{Chakraborty2010}). So we may be witnessing incipient diffusion of Na from olivine margins toward the core, which was more efficient for type I chondrules, closer to batch crystallization \citep{Jacquetetal2015LL,Jacquetetal2012CC}. More direct evidence for diffusion is given by another alkali element, Li, with isotopically lighter and elementally lower compositions in the center of olivine crystals (Neukampf et al., in press), which results from the 5 \% higher diffusivity \citep[e.g.][]{Chakraborty2010} of $^6$Li compared to $^7$Li. Indeed, had diffusion been negligible, the isotopic fractionation (relative to the gaseous reservoir) in olivine cores would be negligible owing to rapid gas-melt equilibration at high temperatures (see section \ref{kinetics}).


\subsection{Glass inclusions}
\label{glass inclusion}

\begin{figure}
\centering
\includegraphics[width=\textwidth]{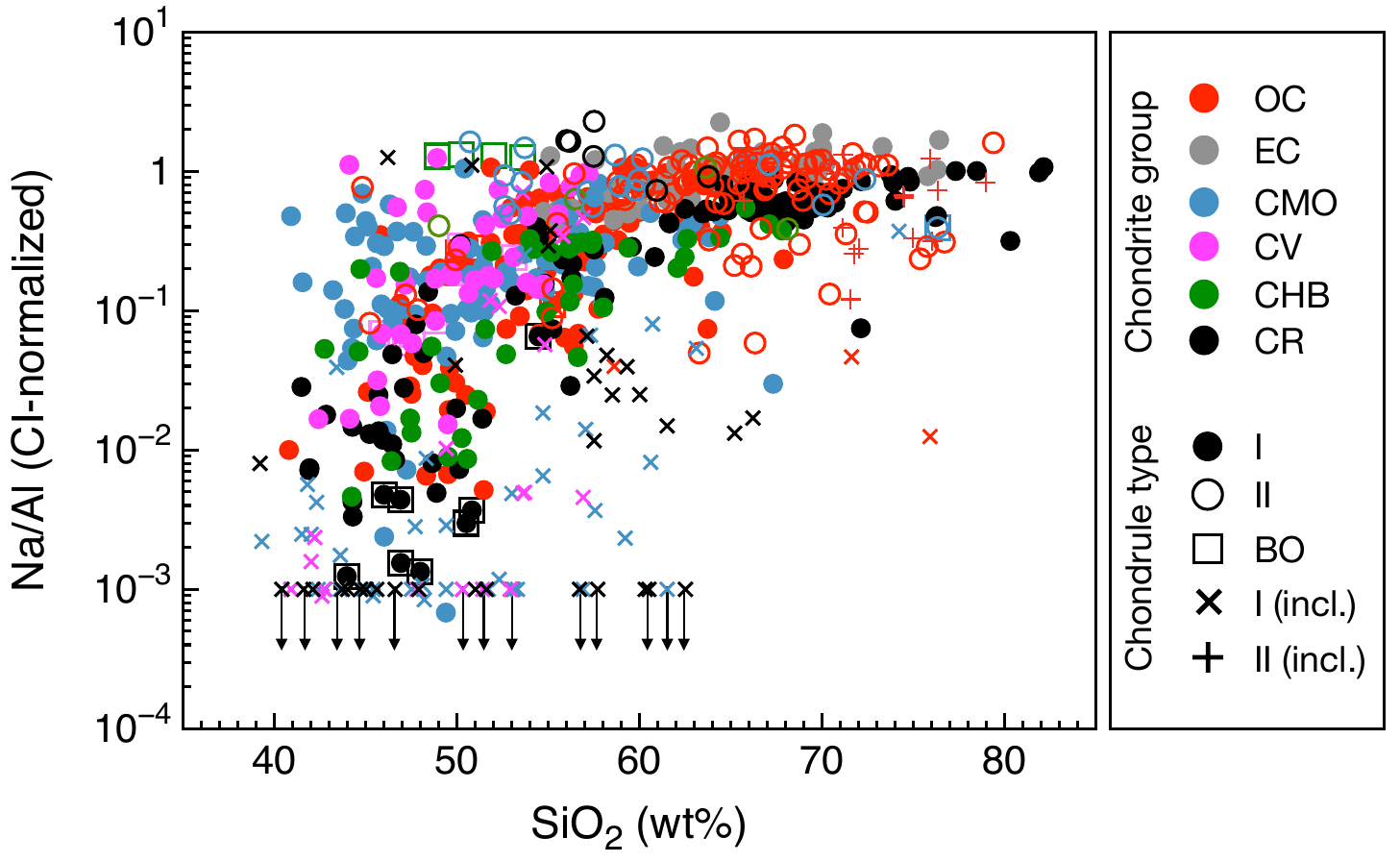}
\caption{Mesostasis/glass Na/Al vs. SiO$_2$ contents for chondrules in different chondrite groups. Colors code the groups and symbol shapes/filling code the petrographic setting. In particular, open circles represent type II chondrules and crosses refer to olivine-hosted glass inclusions ("incl.") from CR \citep{Varelaetal2002}, C3-an Acfer 094 \citep{VarelaKurat2009}, CM \citep{Fuchsetal1973,Desnoyers1980, VarelaKurat2009,Florentinetal2017}, CV \citep{Varelaetal2005,VarelaKurat2009
}, LL \citep{VarelaKurat2009,Hewinsetal2012} chondrites. Values below detection are reported as upper limits at $10^{-3}$. }\label{NaAl_vs_SiO2}
\end{figure}

    The olivine zoning profiles are thus equivocal at best, but olivine-hosted glass inclusions offer a clear message. Indeed, they represent melt trapped earlier than that solidified as the mesostasis interstitial to the phenocrysts \citep[e.g.][]{VarelaKurat2009,Florentinetal2017}. The compositional ratio of choice is Na/Al, for it is insensitive to mineral crystallization after trapping (since both Na and Al are very incompatible elements in olivine and pyroxene) and Al is refractory and thus near closed-system behavior. We can see in Fig. \ref{NaAl_vs_SiO2} 
  that glass inclusions are systematically depleted in Na (often below detection) compared to the mesostasis, even (if to a lesser extent) in type II chondrules.  There is a minor Na-rich glass inclusion population in CR chondrites \citep{Varelaetal2002}, which may have been trapped relatively late, or even be connected in the third dimension to the mesostasis, like the (indeed more Na-rich) "neck inclusions" in the CV chondrite Kaba \citep{Varelaetal2005}. 

  To be sure, \citet{Florentinetal2017} reported that most glass inclusions in Allende are Na-rich but these are likely the result of metasomatism on the parent body. Indeed, although most Na-enriched mesostases are devitrified \citep{IkedaKimura1995, KimuraIkeda1995}, closer observations indicate that Na enrichment preceded devitrification \citep{IkedaKimura1996}. The fact that heating experiments on Allende glass inclusions by \citet{Florentinetal2017} caused no Na loss is certainly no contradiction, since heating was the cause of metasomatism. Chondrites heated below metamorphic grade M0.2 (equivalent to subtype 3.2), which is that of Allende in the \citet{JacquetDoisneau2024} classification, uniformly show mostly Na-depleted glass inclusions
. 

  These evidence that earlier (higher-temperature) melts, and thus chondrules as wholes, were alkali-poorer than the later ones. Melts were thus open both to interaction with the gas and silicate subtraction. This is not unlike the "Primary Liquid Condensation model" of \citet{VarelaKurat2009}, although evidence for solid precursors   \citep[e.g.][]{Krotetal2006,Marrocchietal2018,Marrocchietal2019relicts,Pirallaetal2021,MisawaNakamura1988,JacquetMarrocchi2017,Schneideretal2020}    
suggest chondrules were not \textit{wholly} produced by condensation. The chondritic ratios of refractory lithophile elements in glasses put forward by \citet{VarelaKurat2009} simply reflect their incompatibility in the other phases (so that they represent bulk precursor ratios, as the refractory elements in question hardly evaporated from the chondrules). At any rate, chondrules of both types I and II were open systems in alkalis.

\section{Thermodynamics of alkali condensation}
\label{thermodynamics}

\begin{table}
\caption{Table of the main symbols used in the main text (excluding appendices or footnotes). Symbols may be further subscripted by isotopes ($i$, $j$) or element, or phases ("chd"=chondrule, "g"=gas), not all possibilities of which are indicated.}
\label{table symbols}
\begin{tabular}{c c}
\hline
Symbol & Meaning\\
\hline
$\alpha$ & condensation coefficient\\
$\gamma$ & surface tension of molten chondrule\\
$\gamma_{\rm EO_{0.5}}$ & activity coefficient\\
$\delta^{i/j}E$ & relative deviation of $^i$E/$^j$E to the standard\\
$\eta$ & melt viscosity\\
$\Lambda$ & optical basicity\\
$\phi$ & crystal volume fraction in chondrule\\
$\rho_p$ & chondrule density (as a swarm)\\
$\rho_s$ & internal chondrule density\\
$\tau_{\rm eff}$ & equilibration timescale of the gas with the chondrules\\
$\Omega_E$ & diffusion factor of element E\\ 
$a$ & chondrule radius\\
$E_{\rm chd/g}$ & equilibrium budget ratio (i.e. $f_j/(1-f_j)$) of element $E$\\
$(E/F)_{\rm chd/g}$ & equilibrium ratio of E/F ratios in chondrules and gas\\
Fa & fayalite content in chondrule olivine (=Fe/(Fe+Mg))\\
$f_j$ & condensed fraction of isotope $j$\\
$f_{\rm O_2}$ & oxygen fugacity (dimensionless, i.e. $P_{O_2}/P_0$)\\
$f_{\rm O_2, IW}$ & oxygen fugacity of the Iron-Wüstite (IW) buffer\\
$K_{\rm Na}$ & evaporation reaction constant for Na\\
MDF & mass-dependent fractionation per relative mass difference\\
$m_{\rm Na}$ & mass of Na atom\\
$m_l$ & mean molecular weight in melt\\
$m_p$ & chondrule mass\\
Na$_l$ & mass fraction of Na in melt\\
$N_i$ & total atom number density of isotope $^i$E\\
$n_i$ & number density of gaseous isotopolog $^i$E\\
$n_p$ & chondrule number density\\
$P_{\rm Na}$ & partial pressure of Na\\
$P_0$ & standard pressure (1 bar)\\
$t_{\rm rec}$ & equilibration (recondensation) timescale of a single chondrule with the gas\\
$T_{\rm Fa}$ & numerator in the Arrhenian dependence of $f_{O_2}$ as a function of Fa\\
$T_{\rm Na}$ & numerator in the Arrhenian dependence of $K_{\rm Na}$\\
$T_{\rm IW}$ & numerator in the Arrhenian dependence of $f_{\rm O_2,IW}$\\
$T_{\rm rec}$ & numerator in the Arrhenian dependence of $t_{\rm rec}$, $E_{\rm chd/g}$\\
$v_T$ & mean thermal velocity\\
$x_l$ & melt mass fraction in chondrule\\
$X_{\rm Na}$ & number of Na atoms in the chondrule of interest\\
$X_{\rm buf, Na}$ & value of $X_{\rm Na}$ buffered by the ambient $n_{\rm Na}$\\
$w$ & shape-dependent parameter in the expression of $D_{\rm eff}$\\
\hline
\end{tabular}
\end{table}

  It is time to get more quantitative by modelling the equilibrium between the gas and chondrule melt. We consider a chondrule composed of a silicate liquid (with a mass fraction $x_l$, which may depend on time) and other phases (e.g. silicate crystals, possibly with melt inclusions) which are negligible carriers of the alkali budget and thus not further modelled here (however useful their trace amounts may be to track condensation history as discussed in the previous section). For simplicity, we focus on sodium, but any notation to follow with an "Na" subscript may be adapted for another alkali, and thermodynamic evaluations for them are given in \ref{alkali thermo}. Table \ref{table symbols} recapitulates the main symbols used in the text.  

  We call $K_{\rm Na}$ the equilibrium constant of the reaction 
\begin{equation}
\rm NaO_{0.5}(l)=Na (g)+\frac{1}{4}O_2 (g)
\end{equation}
 Unless otherwise noted, the free enthalpies of chemical reactions were fitted to NIST-JANAF thermodynamical data available online \citep{StullProphet1971}. Between 1200 and 3000 K, a satisfactory fit is:
\begin{equation}
K_{\rm Na}=9.5\times 10^{5}\mathrm{exp}\left(-\frac{T_{\rm Na}}
{T}\right)
\end{equation}
with $T_{\rm Na}=30686$ K (essentially the enthalpy of the above reaction in standard conditions, divided by the Boltzmann constant $k_B$). This is 0.0103 at 1400$^\circ$C (that is 1673 K) which will serve as our reference temperature because most Na solubility experiments conducted or compiled by \citet{Mathieu2009} were at this temperature.

From the action mass law, we can derive:
\begin{equation}
\mathrm{Na}_l=\frac{P_{\rm Na}f_{O_2}^{1/4}m_{\rm Na}}{P_0 K_{\rm Na}\gamma_{\rm NaO_{0.5}}m_l}
\end{equation}
with $P_{\rm Na}$ the partial pressure of Na, $f_{O_2}$ the oxygen fugacity, $P_0=1$bar the standard pressure, $m_{\rm Na}$ the mass of the Na atom, $m_l$ the mean "molecular" (oxide) mass in the melt, $\gamma_{\rm NaO_{0.5}}$ the activity coefficient of NaO$_{0.5}$. The calculation of the activity coefficients is detailed in \ref{appendix_gamma}. Suffice it to say that it is an increasing function of the optical basicity $\Lambda$ which anticorrelates with silica content. In other words, sodium solubility correlates with silica.

  We may reason in terms of the chondrule bulk Na concentration Na$_{\rm chd}=\mathrm{Na}_l x_l$\footnote{This breaks down for Li, with olivine/melt partition coefficients of 0.13-0.35 \citep{SpandlerONeill2009}, in which case we would need to replace $x_l$ by $x_l+\sum_i x_i D_{i/l}$ with $x_i$ the mass fraction of crystalline phase $i$ and $D_{i/l}$ its partition coefficient with the liquid.}. Averaging over all chondrules, and noting $\rho_p$ their local density, we obtain the ratio of Na budgets in the chondrules to that in the gas:
\begin{eqnarray}
\label{Nachd/g}
\mathrm{Na}_{\rm chd/g}&=&\frac{k_B T f_{O_2}^{1/4}\rho_p}{P_0K_{\rm Na}}\langle \frac{x_l}{\gamma_{\rm NaO_{0.5}}m_l}\rangle\\
&=& 2\times 10^{-3}\left(\frac{\langle x_l/(\gamma_{\rm NaO_{0.5}}m_l)\rangle}{\rm 40\: amu^{-1}}\right)\left(\frac{\rho_p}{\rm 10^{-6}\: kg/m^3}\right)\left(\frac{f_{O_2}}{f_{\rm O_2,IW}}\right)^{1/4}\nonumber\\
& & \left(\frac{\rm 1673 K}{T}\right)^{0.49} \mathrm{exp}\left(\left(T_{\rm Na}-\frac{T_{\rm IW}}{4}\right) 
\left(\frac{1}{T}-\frac{1}{\rm 1673\:K}\right) \right)
\end{eqnarray}
where $\langle ... \rangle$ denotes chondrule mass-weighted averaging, $k_B$ the Boltzmann constant, and 
\begin{equation}
f_{\rm O_2, IW}=3\times 10^7 \left(\frac{\rm 1673 K}{T}\right)^{2.03}\mathrm{exp}\left(-\frac{T_{\rm IW}}{T}\right)
\end{equation}
the oxygen fugacity of the iron-wüstite (IW) buffer, with $T_{\rm IW}=66261\:$K, for total pressures much lower than atmospheric \citep{Wolfetal2023}
. This normalization is chosen because $f_{O_2}/f_{\rm O_2,IW}$ weakly varies with temperature in condensation calculations \citep{MokhtariBourdon2026}. The normalizing value for $x_l/(\gamma_{\rm NaO_{0.5}}(1673 \mathrm{K})m_l)$ corresponds to the average of type I OC chondrules (38$\pm$5 amu$^{-1}$ vs. 66$\pm$5 amu$^{-1}$ for type II and 50$\pm$4 and 28$\pm$5 amu$^{-1}$ in EH and CMO respectively; $\pm$ one standard error of the mean, n=28, 56, 36, 28, respectively).

  The compositional evolution of the chondrule during cooling would have had little effect on the right-hand-side of equation \ref{Nachd/g}. Indeed, although crystallization would have reduced $x_l$, $\gamma_{\rm NaO_{0.5}}(1673 \mathrm{K})$ would have decreased too because of increasing silica content in the melt. For chondrules with both bulk and mesostasis data, $x_l/(\gamma_{\rm NaO_{0.5}}(1673 \mathrm{K})m_l)$ of the bulk is higher than the mesostasis by an average ($\pm$standard error of the mean) factor of 4.5$\pm$0.9 (n=28) for type I and 2.9$\pm$0.7 (n=55) for type II chondrules in OC. Yet surely the actual peak temperature composition of the (partially evaporated) chondrules was more refractory than the present-day bulk composition, and would thus have had a higher optical basicity and hence $\gamma_{\rm NaO_{0.5}}(1673 \mathrm{K})$, reducing the above factor. 

  Thus, the explicit temperature dependence would be the main control on Na$_{\rm chd/g}$, and it is a strong one. Indeed Na$_{\rm chd/g}\propto \mathrm{exp}\left(T_{\rm rec, Na}/T\right)$ with  
\begin{equation}
T_{\rm rec, Na}=T_{\rm Na}-\frac{T_{\rm IW}}{4}-(1673\:\mathrm{K})\mathrm{ln}\gamma_{\rm NaO_{0.5}}(1673\:\mathrm{K})=29619\:\mathrm{K}
\end{equation}
 where the numerical evaluation assumes $\gamma_{\rm NaO_{0.5}}(1673 \mathrm{K})=10^{-4}$.  At 1400 $^\circ$C, the "thermal e-folding" $|$d$T/$dlnNa$_{\rm chd/g}|\approx T^2/T_{\rm rec, Na}$ is about 100 K
. So one or a few hundred degrees difference between melt trapping in an olivine crystal and closure of mesostasis would account for the Na-depletion in glass inclusions. Only a competing order-of-magnitude decrease of $\rho_p$ could reverse the effect, as in a plume expansion scenario \citep{Dullemondetal2016}.  

    The strong temperature dependence also means that once the constraint that Na was quantitatively condensed at liquidus \citep{Alexanderetal2008} is eliminated, the density constraints can be alleviated. All normalizations of equation \ref{Nachd/g} equal (with a uniform $\gamma_{\rm NaO_{0.5}}(1673 \mathrm{K})=10^{-4}$), the half-condensation temperature of Na (i.e. where 
Na$_{\rm chd/g}=1$) is: 1120, 1230, 1360 K  for $\rho_p=10^{-7},10^{-6},10^{-5}\:$kg/m$^3$, respectively (this becomes 1030, 1120, 1230 K at IW-4, i.e. $\mathrm{log}_{10}\left(f_{O_2}/f_{\rm O_2, IW}\right)=-4$). These are higher than "canonical" values, e.g. 958 K for a solar gas at $10^{-3}$ bar for \citet{Lodders2003}, which would correspond to more reducing conditions (IW-6; \citet{EbelGrossman2000}) and a relatively low $\rho_p\approx 10^{-7}\:$kg/m$^3$. 

   The higher alkali content of type II chondrules compared to type I chondrules may stem from their higher $f_{O_2}$, but possibly also from higher absolute $\rho_p$. Since $f_{O_2}\propto (O/H-C/H)^{1/2}$ \citep[e.g.][]{Grossmanetal2008}, that is, proportional to the square root of the solid/gas ratio, both might have covaried, e.g. for a constant background gas density. 

\begin{figure}
\centering
\includegraphics[width=\textwidth]{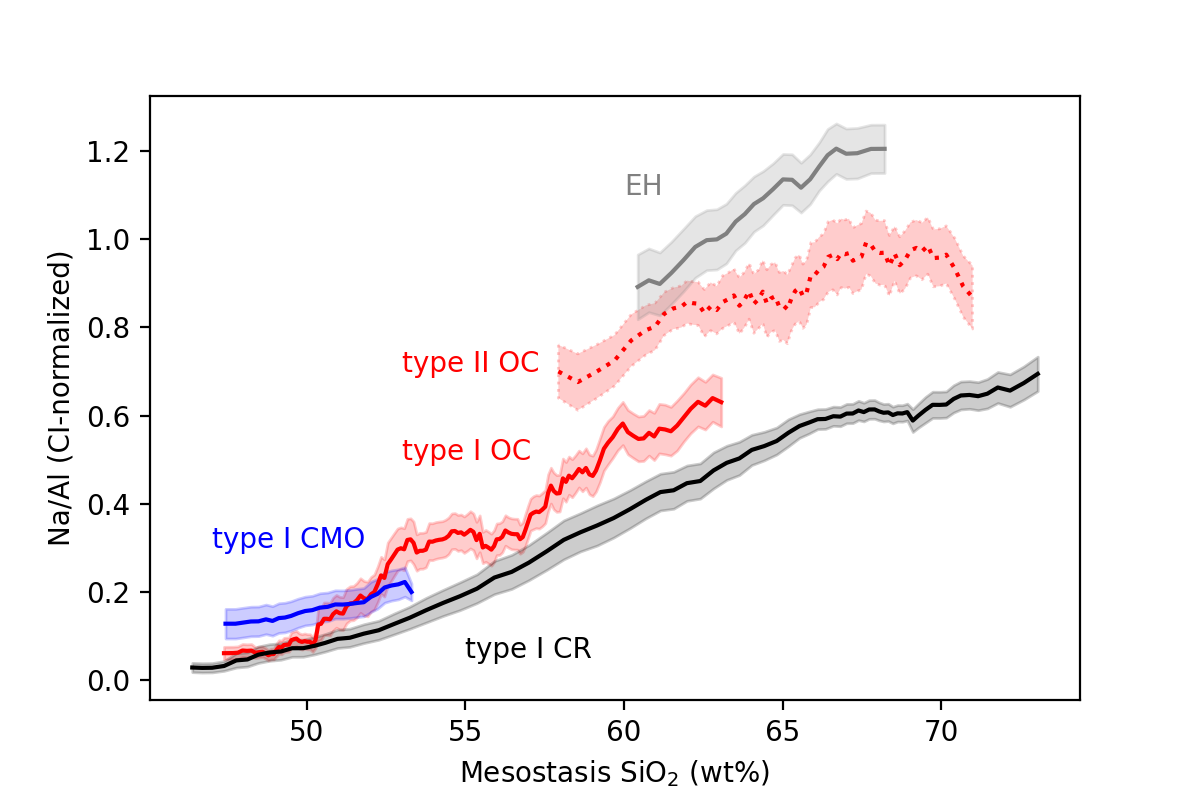}
\caption{Rolling averages of Na/Al (CI-normalized) for mesostasis data (30 averaged at a time) ordered by SiO$_2$ content ($\pm$ one standard error of the mean shaded). As explained in subsection \ref{glass inclusion}, the ratio Na/Al allows to gauge gain of Na, as both Na and Al are incompatible and Al is refractory.
}\label{Na_vs_SiO2}
\end{figure}

\begin{figure}
\centering
\includegraphics[width=\textwidth]{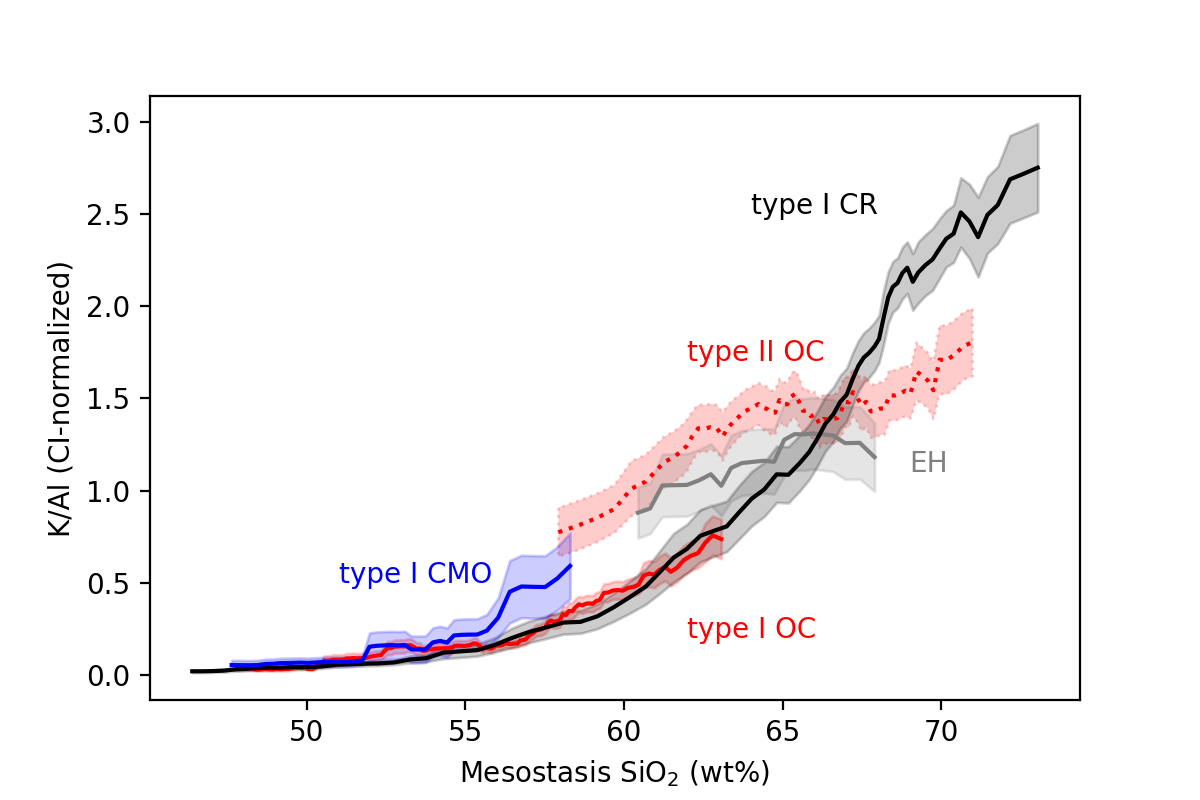}
\caption{Rolling averages of K/Al (CI-normalized) for mesotasis data (30 averaged at a time) ordered by SiO$_2$ content ($\pm$ one standard error of the mean shaded).}\label{K_vs_SiO2}
\end{figure}

\begin{figure}
\centering
\includegraphics[width=\textwidth]{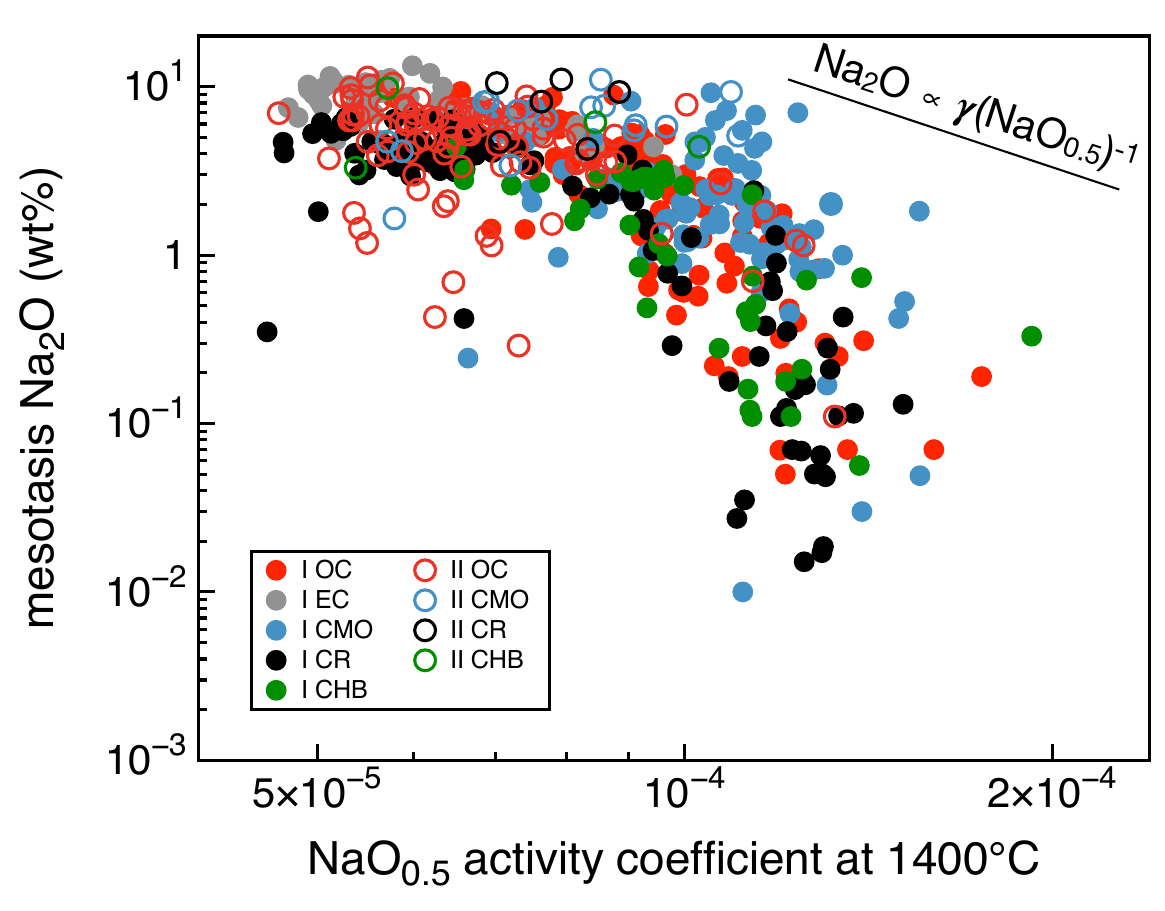}
\caption{Mesostasis Na$_2$O vs. activity coefficient $\gamma_ {\rm NaO_{0.5}}$ in the Henry domain at 1400 °C. The slope of a dependence Na$_2$O$\propto\gamma_ {\rm NaO_{0.5}}^{-1}$ which would apply for constant $T$ and partial pressures of Na and O$_2$ is also shown for comparison.}
\label{Na2O_vs_gamma}
\end{figure}

    The general positive correlation of alkalis with SiO$_2$ (Fig. \ref{Na_vs_SiO2}, \ref{K_vs_SiO2}) of type I chondrule mesotases also supports recondensation. Indeed, alkali solubility is positively correlated with silica content, via the activity coefficients \citep{Liboureletal2003,Mathieu2009}. However, Na varies more steeply than $\gamma_{\rm NaO_{0.5}}^{-1}$ (Fig. \ref{Na2O_vs_gamma}) except when leveling off near chondritic Al-normalized values.  In no chondrite group are we simply witnessing a population of chondrules equilibrated with the same gas, at the same temperature. Since SiO$_2$ contents increased as a result of SiO solubilization \citep{Liboureletal2006}, the chondrules seem to sample an array of different closure temperatures, or time-temperature histories. We thus need to investigate the kinetics of recondensation.

\section{Kinetics of recondensation}
\label{kinetics}

\subsection{Hertz-Knudsen equation}

  Let $X_{\rm Na}$ be the number of Na atoms of the chondrule of interest. Provided the said chondrule is much smaller than the mean molecular mean free path of the gas, it obeys the Hertz-Knudsen equation \citep[e.g.][]{Richter2004,Richteretal2009,Fedkinetal2006}:
\begin{equation}
\label{H-K}
\frac{\mathrm{d}X_{\rm Na}}{\mathrm{d}t}=\pi a^2\left(1-\phi\right)\alpha_{\rm Na}\left(n_{\rm Na}-n_{\rm Na, sat}\right)v_{\rm T,Na}=\frac{X_{\rm buf, Na}-X_{\rm Na}}{t_{\rm rec, Na}}
\end{equation}
with $\alpha_{\rm Na}$ the condensation coefficient (identified with the evaporation coefficient), $n_{\rm Na}$ the concentration of gaseous Na, $n_{\rm Na, sat}$ its saturation value (in equilibrium with the actual melt composition), $v_{\rm T, Na}=\sqrt{8k_B T/(\pi m_{\rm Na})}$ the thermal speed of Na gas molecules. The $(1-\phi)$ factor, with $\phi$ the volume fraction of crystals, is a rough estimate of the fraction of surface of the chondrule occupied by melt (on the one hand, crystals are more concentrated near the surface either as olivine palissades \citep{LibourelPortail2018} or pyroxene rims \citep{Liboureletal2006}, inhibiting exchange \citep{OzawaNagahara2001}, but on the other hand melt may wet them).  We have coined the recondensation timescale
\begin{eqnarray}
\label{trec Na}
t_{\rm rec,Na} &=&\frac{\rho_s a\sqrt{2\pi m_{\rm Na} k_B T} f_{O_2}^{1/4} x_l}{3P_0 K_{\rm Na}\alpha_{\rm Na} m_l \gamma_{\rm NaO_{0.5}}\left(1-\phi\right)}\\
&=& 22\:\mathrm{s} \left(\frac{\rho_s a}{\rm 1 \: kg/m^2}\right)
\left(\frac{0.1}{\alpha_{\rm Na}}\right)\left(\frac{x_l/(\gamma_{\rm NaO_{0.5}}(1673 \mathrm{K})m_l)}{\rm 40\: amu^{-1}}\right)\left(\frac{f_{O_2}}{f_{\rm O_2,IW}}\right)^{1/4}
\nonumber\\
&&\left(\frac{T}{\rm 1673 K}\right)^{0.01}\mathrm{exp}\left(T_{\rm rec,Na}
\left(\frac{1}{T}-\frac{1}{\rm 1673 K}\right)\right) \left(1-\phi\right)^{-1}\nonumber
\end{eqnarray}
with $\rho_s$ the density of the chondrule and $X_{\rm buf, Na}$ 
 the value of $X_{\rm Na}$ buffered by the ambient gas composition. The normalizing value of $\alpha_{\rm Na}$ is inspired from \citet{Fedkinetal2006} who found 0.26 in vacuum and 0.042 under 9 Pa H$_2$. The $f_{O_2}^{1/4}$ dependence was experimentally confirmed by \citet{Tsuchiyamaetal1981}.

$t_{\rm rec, Na}$, which is also an evaporation timescale, is short for typical liquidus temperatures. We calculate 0.3 s 
at 1960 K and IW-3 for our usual normalizations (with $\gamma_{\rm NaO_{0.5}}(1673 \mathrm{K})=10^{-4}$
), comparable to the 0.5 s convergence timescale quoted by \citet{FedkinGrossman2013} for their type I model chondrule composition. No precursor alkali content should thus survive the heating. Indeed:
\begin{eqnarray}
\label{XNa solved}
X_{\rm Na} &=& X_{\rm buf,Na}-\int_0^t\mathrm{exp}\left(-\int_{t'}^t\frac{\mathrm{d}t''}{t_{\rm rec,Na}}\right)\frac{\mathrm{d}X_{\rm buf,Na}}{\mathrm{d}t}\mathrm{d}t'\nonumber\\
&+&\left(X_{\rm Na}(0)-X_{\rm buf,Na}(0)\right)\mathrm{exp}\left(-\int_0^t\frac{\mathrm{d}t'}{t_{\rm rec,Na}}\right)\nonumber\\
&\approx& X_{\rm buf, Na}-t_{\rm rec,Na}\frac{\mathrm{d}X_{\rm buf,Na}}{\mathrm{d}t}
\end{eqnarray}
where the approximation assumes that $t_{\rm rec, Na}$ is shorter than both the elapsed time (henceforth the \textit{limit of forgotten beginnings}) and the timescale of variation of $X_{\rm buf, Na}$--or more precisely for the latter condition, that:
\begin{equation}
\label{near-eq condition}
\frac{\rm d}{\mathrm{d}t}\left(t_{\rm rec, Na}\frac{\mathrm{d}X_{\rm buf, Na}}{\mathrm{d}t}\right)/\frac{\mathrm{d}X_{\rm buf, Na}}{\mathrm{d}t}\ll 1
\end{equation}
However, as the
 temperature decreases, the lag time $t_{\rm rec}$ increases and ultimately overtakes the cooling timescale
, effectively stopping the alkali influx.

\subsection{Isotopic fractionation}
\label{MDF subsection}

The recondensation timescale $t_{\rm rec}$ being shorter for lighter isotopologs (for alkalis other than Na which are multi-isotopic in nature), we would expect an excess of lighter isotopes in the chondrule. Indeed, by plugging in the relevant Hertz-Knudsen equations, the isotopic ratio for two isotopes $i$ and $j$ of an element $E$ evolves as:
\begin{eqnarray}
\label{d ratio/dt}
\frac{\mathrm{d}}{\mathrm{d}t}\left(\frac{X_i}{X_j}\right)=\frac{1+\sqrt{\frac{m_i}{m_j}}\alpha^{i/j}_{\rm chd-g}\left(\frac{X_{\rm buf, j}}{X_j}-1\right)}{t_{\rm rec,i}}\nonumber\\
\left[\frac{\alpha^{i/j}_{\rm chd-g}n_i/n_j}{1+\left(\sqrt{\frac{m_i}{m_j}}\alpha^{i/j}_{\rm chd-g}-1\right)\left(1-\frac{X_j}{X_{\rm buf,j}}\right)}-\frac{X_i}{X_j}\right]
\end{eqnarray}
with $\alpha^{i/j}_{\rm chd-g}$ the chondrule/gas (equilibrium) fractionation coefficient, where we have ignored any mass-dependence of the condensation coefficients \citep{Zhangetal2021}. Hence, $X_i/X_j$ should converge rapidly toward the first term between the brackets (so long as inequation \ref{near-eq condition} applies). In terms of the relative deviations of relative deviation of the $^i$E/$^j$E ratio to a terrestrial standard, we would have:
\begin{eqnarray}
\label{delta chd-g}
\delta^{i/j}E_{\rm chd}  -\delta^{i/j}E_{\rm g}&\approx&\frac{X_i n_j}{X_j n_i}-1\nonumber\\
&\approx& \alpha^{i/j}_{\rm chd-g}-1 + \left(\sqrt{\frac{m_i}{m_j}}-1\right)\left(\frac{X_j}{X_{\rm buf,j}}-1\right)\nonumber\\
&\approx& -\left(\sqrt{\frac{m_i}{m_j}}-1\right)t_{\rm rec,j}\frac{\mathrm{dln}X_{\rm buf,j}}{\mathrm{d}t}
\end{eqnarray}
Since $X_j/X_{\rm buf,j}=n_{\rm sat,j}/n_j$, the second form amounts to equation 31 of \citet{Richter2004} when equilibrium fractionation is neglected ($\alpha^{i/j}_{\rm chd-g}\approx 1$), as done in the third form\footnote{From the values quoted by \citet{Zhangetal2021}, the correction to the MDF (defined later), $(1-f_j)(\alpha^{i/j}_{\rm chd-g}-1)/(m_i/m_j-1)$, would be a positive MDF shift of $0.001-0.002$ for both K and Rb at 1400 °C (with a dependence $\propto 1/T^2$) in the limit $f_j\ll 1$. This is subordinate compared to the MDF values tabulated in \ref{table MDF}, which are dominated by the (negative) kinetic contribution.} .

Assuming the  whole chondrule-forming region (CFR) comprises a population of identical-composition 
 chondrules, in addition to the gas, with overall concentration (condensed+gaseous) $N_i$ and $N_j$ for the isotopes in question, we have:
\begin{equation}
\delta^{i/j}E_{\rm chd}-\delta^{i/j}E_{\rm CFR}=-\left(1-f_j\right)\left(\sqrt{\frac{m_i}{m_j}}-1\right)\langle t_{\rm rec,j}\rangle\frac{\mathrm{dln}}{\mathrm{d}t}\left(\frac{N_j/n_p}{\mathrm{E}_{\rm chd/g}^{-1}+1}\right) 
\end{equation}
with $f_j$ the recondensed fraction of isotope $j$ (essentially that of the element $E$ as a whole) and
$n_p$ the number density of chondrules. Assuming $N_j/n_p$ to be constant (which would still allow for expansion of the coupled gas+chondrule cloud), and using
\begin{equation}
\frac{t_{\rm rec,j}}{E_{\rm chd/g}}=\frac{1-f_j}{f_j}t_{\rm rec,j}=\frac{4\rho_sa}{3\rho_pv_{T,j}\alpha_j\left(1-\phi\right)}\equiv\tau_{\rm eff,j},
\end{equation}
we can express the mass-dependent fractionation per relative mass difference MDF as:
\begin{eqnarray}
\label{MDF equation}
\mathrm{MDF}&\equiv& \frac{\delta^{i/j}E_{\rm chd}-\delta^{i/j}E_{\rm CFR}}{m_i/m_j-1}\\
&=& -\frac{\tau_{\rm eff,j}}{2}
f_j\left(1-f_j\right)\frac{\rm dlnE_{chd/g}}{\mathrm{d}t}\nonumber\\
&=& -0.02 \left(\frac{f_j\left(1-f_j\right)}{10^{-1}}\right)\left(\frac{\langle \rho_s a\rangle}{\rm 1\: kg/m^2}\right)\left(\frac{\rm 10^{-6}\:kg/m^3}{\rho_p}\right)
\left(\frac{0.1}{\alpha_j}\right)\left(1-\phi\right)^{-1}\nonumber\\
 &&\left(\frac{m_j}{\rm 39 \: amu}\right)^{1/2}\left(\frac{\rm 1500 K}{T}\right)^{1/2}  \left(\frac{\rm 100 K}{|\mathrm{d}T/\mathrm{dlnE_{chd/g}}|}\right)\left(\frac{|dT/dt|}{\rm 10 \:K/h}\right)\nonumber
\end{eqnarray}
The parabolic $f_j\left(1-f_j\right)$ dependence reproduces those calculated by \citet{Nieetal2021} for constant cooling rates (if skewed past $f_j=1/2$ by the increase of $|\mathrm{dln\mathrm{E}_{chd/g}}/\mathrm{d}T|$). This comes about because for $f_j\ll 1$, equilibration is efficient and kinetic mass fractionation thus minimal, while for $f_j$ approaching 1, the chondrules have essentially recovered all the $E$ budget of the reservoir, erasing any isotopic fractionation relative to it. As in \citet{Richter2004}, we see that the MDF has a small absolute value so long as the cooling timescale ($\sim \rm dt/dlnE_{chd/g}$) is long compared to the gas equilibration timescale $\tau_{\rm eff}$.

  Carbonaceous chondrite chondrule MDF values are calculated for alkali and other moderately volatile elements in table \ref{table MDF}. We assume that the (mass-dependent) isotopic composition of the reservoir prior to chondrule formation was that of CI chondrites. This neglects nucleosynthetic contributions to mass-dependent isotopic variations. The K nucleosynthetic anomaly variations in $\epsilon^{40}\mathrm{K}\approx\delta^{40/39}\mathrm{K}-0.5\delta^{41/39}\mathrm{K}$ are of order 0.1 \textperthousand$\:$ \citep{Nieetal2023},  but, if indeed ascribable to type II supernovae, are essentially due to $^{40}$K: for the yields adopted by \citet{Nieetal2023}, fresh supernova ejecta should have $\delta^{41/39}\mathrm{K}/\epsilon^{40}\mathrm{K}=10^{-2}$ so their admixtures would not appreciably affect the $\delta^{41/39}$K. Likewise, mass-independent nucleosynthetic anomalies in Zn vary by less than 0.01 \textperthousand   $\:$ among bulk CCs \citep{Savageetal2022} and thus suggest similar inherited variations for mass-dependent ratios (prior to evaporation), negligible compared to the fraction of permil effects observed. The Li isotopic variations are also limited among bulk carbonaceous (or unequilibrated ordinary) chondrites \citep{Seitzetal2007}. 

  The calculated MDF have similar orders of magnitude for the different moderately volatile elements. Indeed, so long as we deal with trace elements dissolving in some phase, the mathematical treatment above should be similar \citep[e.g.][]{Loddersetal2025condensation}, although relevant gas fugacities (other than that of the element-containing species, e.g. O$_2$ here) may differ. The temperature dependence the chondrule/gas budget ratios should be largely dictated by the enthalpy of the condensation reactions, which have similar orders of magnitude (hundreds of kJ/mol; e.g. table 5 of \citet{Sossietal2019MOVE}; Table \ref{table TE Trec} shows $T_{\rm rec}$ only ranging from 3$\times 10^4$ to 5$\times 10^4\:$ K for alkalis). Any variation of the chondrule density would affect all elements the same way, and the molecular mass (tens of amu) and condensation coefficients (0.017-0.19 for major lithophile elements; \citet{Fedkinetal2006}) should be also of the same orders of magnitude. We would need to know those condensation coefficients for the specific elements to make more quantitative comparisons. Elements diffusing slower than alkalis may be affected by mass-dependent diffusivities (see \ref{diffusion}), hence perhaps the more negative MDF for Ge (whose tetravalence makes it one of the slowest diffusers reviewed by \citet{Zhangetal2010}).

\begin{figure}
\centering
\includegraphics[width=\textwidth]{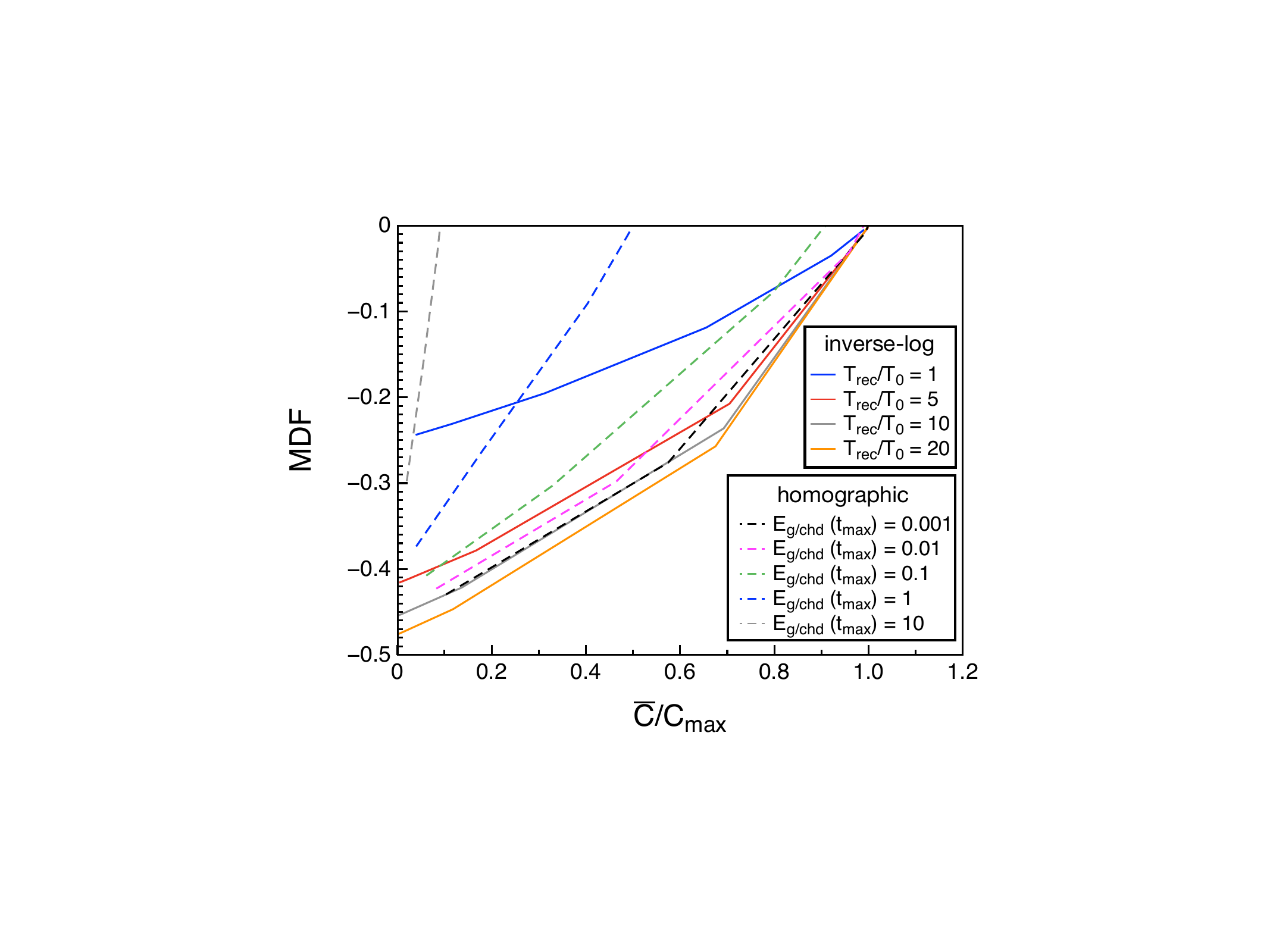}
\caption{MDF as a function of condensed fraction $\overline{C}/C_{\rm max}$ for two different cooling laws: a slow cooling interrupted at $t_{\rm max}$ (\textit{homographic}, dashed line) and an accelerating cooling (\textit{inverse-log}, continuous line). The curves are drawn for different values of parameters (E$_{\rm g/chd}(t_{\rm max})$ and $T_{\rm rec}/T_0$, respectively, see \ref{appendix MDF}). For low condensed fractions, MDF is usually close to -0.5 unless the condensed fraction was dictated by the quench (at $t_{\rm max}$, for the homographic cooling, if E$_{\rm g/chd}(t_{\rm max})$ is high
.}
\label{MDF_vs_C}
\end{figure}

  However, theoretically, these low |MDF| values hold in the limit of near-equilibration between gas and melt. As the temperature decreases, $t_{\rm rec}\mathrm{dln}X_{\rm buf,j}/\mathrm{d}t$ should become of order unity, and the MDF should be close to -0.5 (according to equation \ref{delta chd-g}), as the chondrule becomes closed to the alkali influx (unless the chondrules have recovered most of them, which is not the case of type I chondrules, dominant in CCs). This is illustrated in Fig. \ref{MDF_vs_C} following the calculations of \ref{appendix MDF} for a nonexpanding system. The calculations of \citet{OzawaNagahara2001}, assuming constant cooling rates, likewise invariably found final MDF near -0.3. 
One way out is that the increase of $t_{\rm rec,j}$ suddenly (that is, over a timescale $\ll t_{\rm rec,j}$) accelerated, thus prematurely violating inequation \ref{near-eq condition} (because of the second derivative of $X_{\rm buf,j}=n_j\pi a^2\alpha_j\left(1-\phi\right)v_{T,j}t_{\rm rec,j}$
). This would cut short any further (near-)equilibration, and the cooling rate in equation \ref{MDF equation} would then refer to that before this cooling acceleration.  Such a quenching is supported by the preservation of glassy mesostasis (requiring  cooling rates in excess of 360-1800 K/h or 1-10 K/h for basaltic and rhyolitic melt compositions, respectively, near plausible glass transition temperatures of 680-880$^\circ$C; \citet{Jonesetal2018}), the frequent monoclinic structure of low-Ca pyroxene \citep[requiring 250-10000 K/h around 1000$^\circ$C; e.g.][]{Soulie2014}, or the low density of dislocations in olivine (suggesting quenching above 1000$^\circ$C; \citet{Jonesetal2018}). These indeed contrast with evidence for slower cooling rates at higher temperatures for type I chondrules, e.g. from Cu-Ga zoning in metal grains (0.5-50 K/h at 1200$\pm 100^\circ$C; \citet{Humayun2012,Chaumardetal2018}), clinopyroxene exsolution lamellae (0.1-50 K/h at 1200-1400$^\circ$C; \citet{Weinbruchetal2001}) or mesostasis-olivine partitioning equilibration of trace elements ($\lesssim$10 K/h; \citet{Jacquetetal2012CC}). So (type I) chondrules do seem to have undergone a quench \citep[e.g.][]{Jacquetetal2012CC,Villeneuveetal2015,Liboureletal2023}, perhaps upon their exit from their hot melting region.

\begin{table}
\caption{Mass-dependent fractionations for different elements arranged in order of decreasing CI-normalized abundances ("chd/CI") in CC chondrules (generally the extrapolated "non-matrix component", except for Li where individually measured Allende chondrules could be averaged). $\delta^{i/j}E$ is 
expressed in \textperthousand $\equiv 10^{-3}$
, with 2$\sigma$ errors given. The MDF are calculated assuming 
 a CI isotopic composition of the pre-chondrule formation reservoir. 
}
\label{table MDF}
\begin{tabular}{c c c c c c c c c}
\hline
$E$ & chd/CI & $i$ & $j$ & $\delta^{i/j}E_{\rm CI}$ (\textperthousand) & $\delta^{i/j}E_{\rm chd}$ (\textperthousand) & MDF & Reference\\
\hline
Li & 0.29 & 7 & 6 & 3.6$\pm$0.7 & -0.3$\pm$0.8 & -0.02$\pm$0.01 & \citet{Seitzetal2007,Seitzetal2012}\\
K & 0.48 & 41 & 39 & -0.07$\pm$0.03 & -0.33$\pm$0.12 & -0.005$\pm$0.002 & \citet{Nieetal2021}\\
Rb & 0.43 & 87 & 85 & 0.17$\pm$0.02 & 0.04$\pm$0.05 & -0.006$\pm$0.002 & \citet{Nieetal2021}\\
Ge & 0.25 & 74 & 70 & 1.00$\pm$0.04 & -2.62$^{+0.87}_{-1.13}$ & -0.06$\pm$0.02 & \citet{Woelferetal2025}\\
Te & 0.15 & 130 & 125 & 0.86$\pm$0.19 & -0.47$\pm$0.30 & -0.03$\pm$0.01 & \citet{Mortonetal2024}\\
Zn & 0.11 & 66 & 64 & 0.45$\pm$0.10 & -0.10$\pm$0.03 & -0.018$\pm$0.003 & \citet{Mortonetal2024}\\
Cd & 0.11 & 114 & 110 & 0.38$\pm$0.17 & -0.19$\pm$0.27 & -0.02$\pm$0.01 & \citet{Mortonetal2024}\\
\hline
\end{tabular}
\end{table}

\subsection{Alkali-zoned chondrules}

\begin{figure}
\centering
\includegraphics[width=\textwidth]{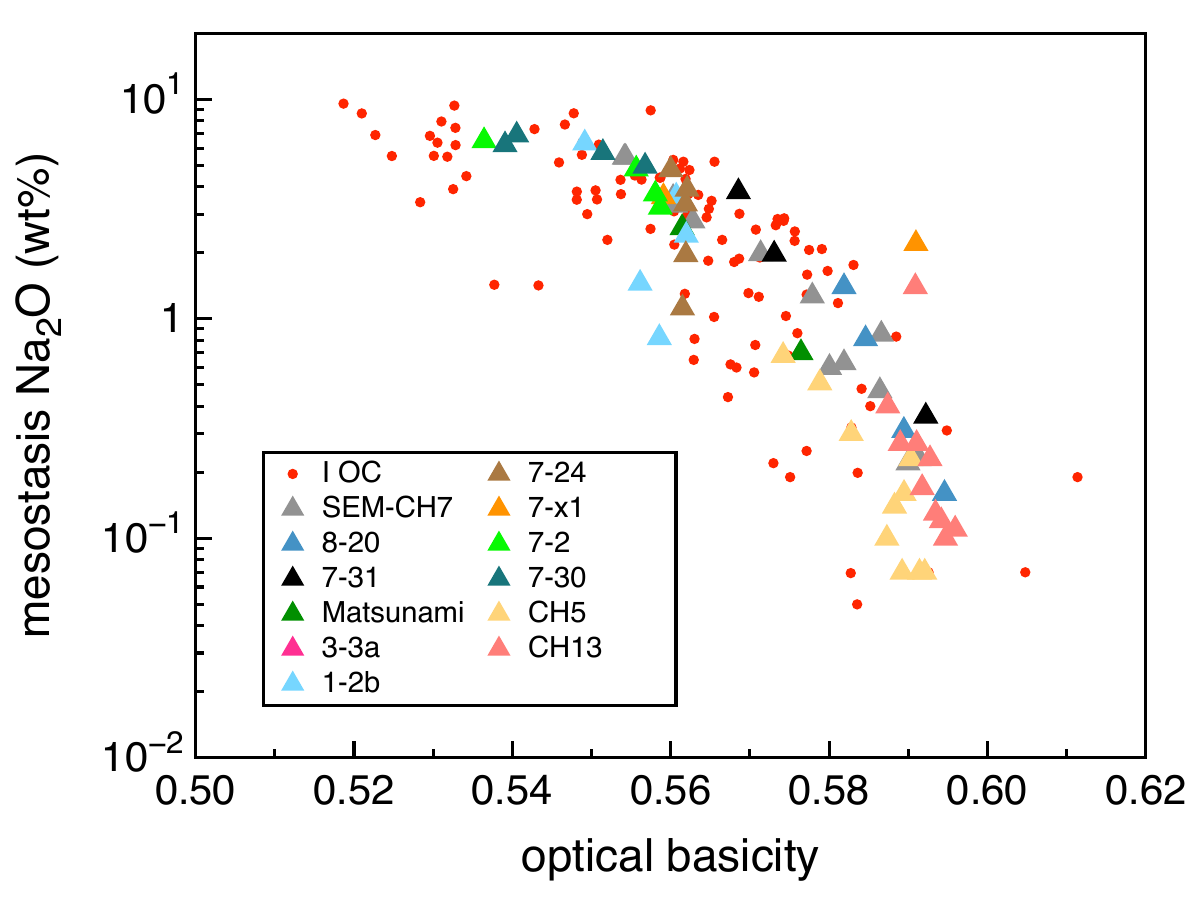}
\caption{Mesostasis Na$_2$O versus optical basicity $\Lambda$ for different alkali-zoned type I chondrules in Semarkona: "Matsunami" \citep{Matsunamietal1993},  SEM-CH7 \citep{Liboureletal2003,Mathieu2009}, CH5 and CH13 \citet{Nagaharaetal2008} all others from \citet{Grossmanetal2002}. They are compared to mesostasis compositions in the general population of type I chondrules in ordinary chondrites.}
\label{alkali_zoned}
\end{figure}

  How are we to understand alkali-zoned chondrules, with alkalis increasing toward their rims \citep{Matsunamietal1993,Grossmanetal2002,Liboureletal2003,Nagaharaetal2008}? In many cases, Na$_2$O is positively correlated with SiO$_2$ (both increasing outward), and thus anticorrelated with optical basicity, as in the general population of chondrules 
 (Fig. \ref{alkali_zoned}). Yet, as in the latter, this is a stronger dependence than a uniform NaO$_{0.5}$ activity throughout the chondrule interior would impose. This is at variance with the conclusion of \ref{diffusion} that diffusion transport was efficient when chondrules closed themselves to alkali recondensation. \citet{Grossmanetal2002} also argued against diffusion because of the near parallel profiles for elements with different diffusivities. 

  These objections assume that the chondrules were always single, isolated entities (apart from gas-melt exchange). However, \citet{Jacquet2021} made the case that most chondrules have undergone collisions while partially molten, producing visible compound chondrules when the relaxation to sphericity was longer than cooling. The coarse-grained (igneous) rims around many chondrules were ascribed to the aggregation of smaller droplets (e.g. microchondrules) around bigger objects near the end of cooling (rather than the secondary melting of dust mantles accreted after solidification; \citet{Jacquet2021}). Since those smaller droplets would have equilibrated with the gas at lower temperatures (all else being equal), or have arisen by collisional disruption preferentially in denser areas, they would tend to be SiO$_2$- and alkali-enriched.

  There would have been a temperature below which diffusion would not have been able to homogenize the chondrules. It would be comparable to the temperature at which compound chondrules made of subequal components can no longer relax to sphericity (after the quench), 1220-1360 K in the estimates of \citet{Jacquet2021} (much lower than typical liquidus temperatures, e.g. 1500-2200 K according to \citet{HewinsRadomsky1990}). Indeed, by taking into account the effect of crystals on diffusivity and the $\pi^2$ factor in the definition of the diffusion timescale $t_{\rm diff}$ (\ref{diffusion}), the ratio of the latter to the sphericization timescale $t_{\rm sph}$ is corrected from equation 5 of \citet{Jacquet2021} as:
\begin{eqnarray}
\frac{t_{\rm diff}}{t_{\rm sph}}&=&\frac{a\gamma \left(1-\phi\right)^{3/2}\left(1+\phi/w\right)}{\Omega_E T\pi^2}\\
&=& 4\left(\frac{a}{\rm 0.3\: mm}\right)\left(\frac{1-\phi}{0.1}\right)^{3/2}\left(1+\phi/w\right)\left(\frac{\rm 1000\: K}{T}\right)\left(\frac{\rm 10^{-10} N/K}{\Omega_E}\right)\nonumber
\end{eqnarray}
with $\gamma=0.36$ N/m the surface tension of the melt, $\phi$ the crystal volume fraction, $w$ a parameter defined in \ref{diffusion} and $\Omega_E$ the "diffusion factor" of \citet{Mungall2002} for the element E of interest. The latter's normalization value is a suitable upper bound for Na in basalt and andesite \citep{Mungall2002}. If Na (activity) cannot be homogenized in the melt, the other elements cannot either, and thus their zonings will all essentially follow the boundary between the accreted components, hence the parallelism observed by \citet{Grossmanetal2002}. \citet{Grossmanetal2002} themselves entertained late addition of material for some zoned chondrules.


  This collisional scenario also accounts for the rarer cases where Na zoning seems decorrelated from optical basicity  (Fig. \ref{alkali_zoned}). Indeed, droplet composition would not simply be a function of size, but also of precursor nonvolatile element abundances.  If the biggest component droplet was e.g. less refractory than average, its melt may have a relatively low optical basicity comparable to that of smaller droplets accreted around it. This would thwart the usual outward decreasing trend of the optical basicity, although alkalis would still increase outward. The diversity of colliding droplets could be further compounded by mixing droplets of different time-temperature histories, e.g. as a function of provenance depth within the chondrule-forming cloud.

\section{Consequences on the astrophysical context of chondrule formation}
\label{context}

\subsection{Chondrule-forming densities}
\label{CFR densities}

  \begin{figure}
\centering
\includegraphics[width=\textwidth]{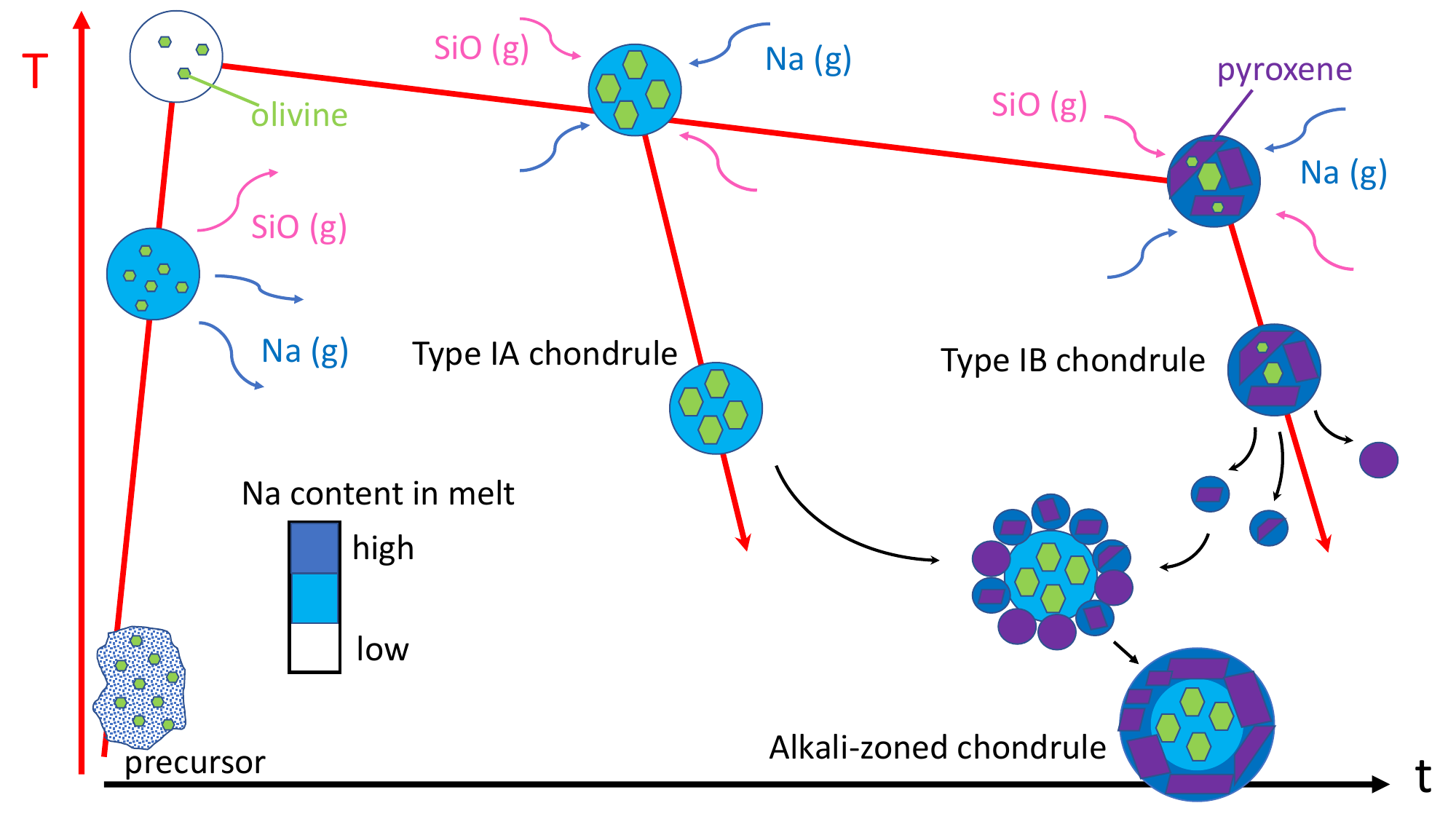}
\caption{Sketch of our alkali (here, Na) recondensation scenario after the melting and partial evaporation of the precursor (here for type I chondrules). For a given smooth default cooling curve, we show two quench trajectories, one at high temperature leading to a Na-poor type IA (olivine-rich) chondrule, and another at lower temperature leading to a Na-rich type IB (pyroxene-rich) chondrule. Of course all intermediate possibilities (e.g. IAB chondrules) exist. A potential agregation between high- and low-temperature droplets (here due to disruption of larger chondrules after quenching, but the disruption could have occurred before, or the microdroplets have been small from the outset) leading to an alkali-zoned chondrule is also depicted.}
\label{schema_Na}
\end{figure}

  Our scenario of alkali recondensation is depicted in Fig. \ref{schema_Na}. We have hitherto refrained from commenting on implied densities. With the \citet{Alexanderetal2008} contention that alkali contents were fixed at liquidus set aside, we certainly no longer need values as high as their calculated $0.002-9.08$ kg/m$^3$. If we knew what equilibration temperature to substitute, we could invert equation \ref{Nachd/g} as:
\begin{eqnarray}
\label{rhop from Nachd/g}
\rho_p &=&\frac{P_0K_{\rm Na}\mathrm{Na}_{\rm chd/g}}{k_BTf_{O_2}^{1/4}\langle x_l/(\gamma_{\rm NaO_{0.5}}m_l) \rangle}\\
&=& 1.3\times 10^{-4}\:\mathrm{kg/m^3}
\left(\frac{\rm Na_{chd/g}}{0.1}\right)
\left(\frac{\rm 1\:mol\%}{\rm Fa}\right)^{1/2}
\left(\frac{\rm 40\:amu^{-1}}{\langle x_l/(\gamma_{\rm NaO_{0.5}}m_l) \rangle}\right)\nonumber\\
&&\left(\frac{\rm 1673\:K}{T}\right)
\mathrm{exp}\left(\left(T_{\rm Na}-\frac{T_{\rm Fa}}{4}\right)\left(\frac{1}{\rm 1673\:K}-\frac{1}{T} \right)\right)
\end{eqnarray}
where the fayalite content Fa of the chondrule-hosted olivine enters via the following relationship \citep{Johnson1986}, in the presence of enstatite\footnote{This thus fails for IIA chondrules in CCs.}:
\begin{equation}
f_{O_2}=8.7\times 10^7\mathrm{Fa}^2\mathrm{exp}\left(-\frac{T_{\rm Fa}}{T}\right)\approx 0.3\mathrm{Fa}^2f_{\rm O_2,IW}
\end{equation}
with $T_{\rm Fa}=$66623 K, and Na$_{\rm chd/g}=f_{\rm Na}/(1-f_{\rm Na})$ may be approximated by equating $f_{\rm Na}$ with the average Al-normalized abundance of Na. 

   In principle, the closure temperature would correspond to a $t_{\rm rec}=\tau_{\rm eff}E_{\rm chd/g}$ reaching some fraction of the timescale of variation of the temperature $\propto \left(dT/dt\right)^{-1}$ so that $E_{\rm chd/g}
\propto \rho_p |dT/dt|^{-1}$. Yet this fraction depends on the shape of the cooling curve, which may not be guessed ab initio, especially given the conjectured quenching. 

  In fact, the most robust constraint is given by the limited isotopic fractionation of alkalis (and other moderately volatile elements)
. Inverting equation \ref{MDF equation}, we obtain:
\begin{eqnarray}
\frac{\rho_p}{|\mathrm{d}T/\mathrm{d}t|}= 
10^{-6}\:\mathrm{kg.m^{-3}.K^{-1}.h} \left(\frac{-0.005}{\rm MDF}\right) \left(\frac{f_j\left(1-f_j\right)}{0.2}\right)\left(\frac{\langle \rho_s a\rangle}{\rm 1\: kg/m^2}\right)
\left(\frac{0.1}{\alpha_j}\right)\nonumber\\
\left(\frac{m_j}{\rm 39 \: amu}\right)^{1/2}\left(\frac{\rm 1500\:K}{T}\right)^{5/2}  \left(\frac{T_{\rm rec}}{3\times 10^4\:\mathrm{K}}\right)\left(1-\phi\right)^{-1}\nonumber\\
\left(1+\frac{T}{T_{\rm rec}}\frac{\rm d}{\rm dlnT}\mathrm{ln}\left(\rho_px_l(f_{O_2}/f_{O_2,IW})^{1/4}/m_l\right)+\frac{1}{T_{\rm rec}}\frac{\rm d}{\rm dlnT}\left(T\mathrm{ln}\gamma_{EO_{0.5}}\right) \right) 
\end{eqnarray}
The explicit temperature dependence does not produce more than a factor of two uncertainty, the $f_{O_2}$ dependence has largely disappeared, though the most important effect is incorporated in the $T_{\rm rec}$, where the exact modelling of the activity coefficient is likewise of subordinate importance.
The best constraint would be from potassium, as it should be indeed free from diffusion effect in melt (\ref{diffusion}) or in crystals (because of its incompatibility, unlike Li) 
and the normalizations above are adapted to it. Indeed, type I CMO chondrule mesostases in our compilation have an average CI-normalized K/Al ratio (essentially our $f_j$, since the mesostasis ratio is the bulk ratio) of 0.19 ($\pm$0.03, n=75).
 If the cooling rate is assumed to be no smaller than the lowest literature estimates discussed previously ($\sim 0.1$K/h; \citet{Jonesetal2018}), this indicates $\rho_p\gtrsim 10^{-7}\:$kg/m$^3$. 

  Interestingly, the Rb/K ratios of the alkali-poorest chondrules in Allende are up to twice the chondritic value \citep{Matsudaetal1990}. This suggests that (Rb/K)$_{\rm chd/g}$ was $\sim 2$, which from \ref{alkali thermo} would indicate a quenching temperature around 1250 K. So if this, after all, is a suitable estimate, equation \ref{rhop from Nachd/g} would yield $\rho_p=6\times 10^{-6}\:$kg/m$^3$ (for $\gamma_{\rm Na_{0.5}}(1673\:\mathrm{K})=10^{-4}$).

 \begin{figure}
\centering
\includegraphics[width=\textwidth]{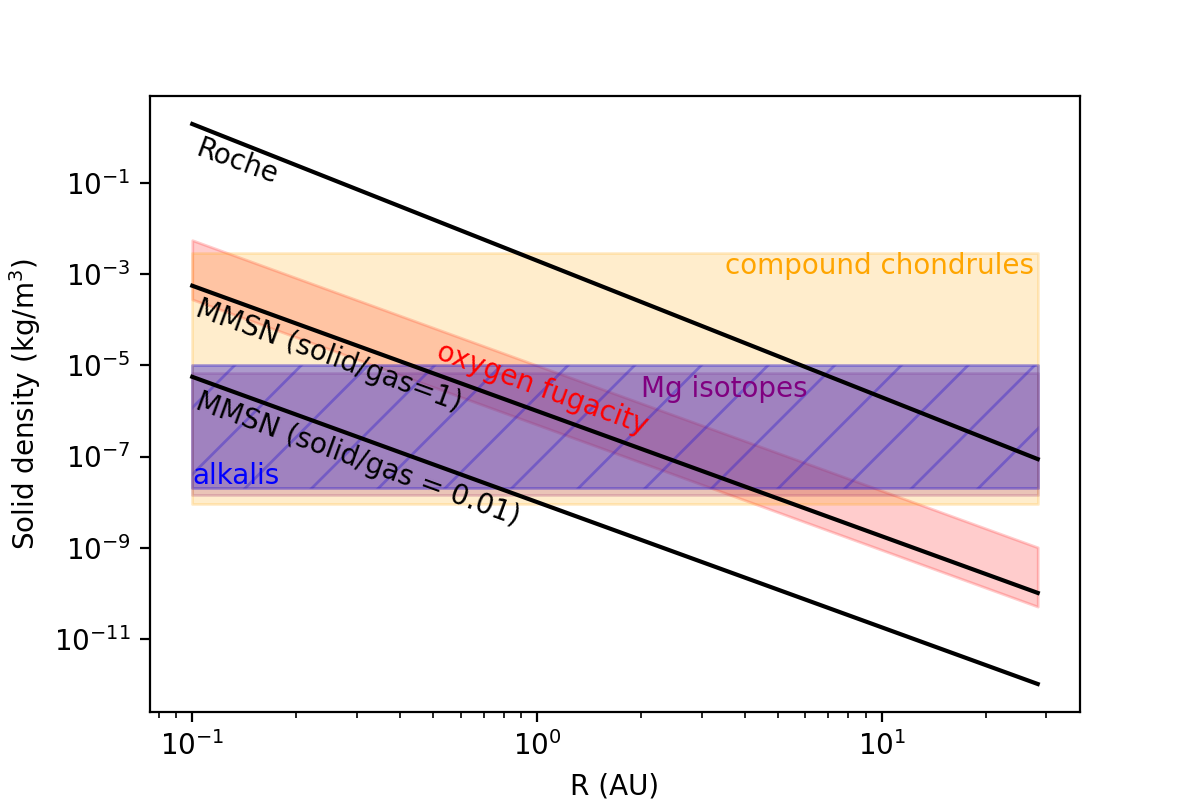}
\caption{Various constraints from this work and the literature (mutatis mutandis) on chondrule-forming solid densities, as a function of heliocentric distance $R$, modified after \citet{Jacquetetal2024}, assuming $a=0.3$ mm. The alkali constraint is derived from this work, chiefly the K isotopic fractionation, assuming no expansion and a (pre-quench) cooling rate comprised between 0.1 and 50 K/h (see discussion in \ref{MDF subsection}). The nearly identical Mg isotopic constraint is from \citet{CuzziAlexander2006}, with their heating time $2\times 10^4$ s taken as a minimum, and the upper bound differing by the same factor as for the cooling rates selected above for the alkalis. The compound chondrule frequency constraint is taken from \citet{Jacquet2021}, with a (post-quench) collision time of 1 h, N$_{\rm coag}$ comprised between 0.1 and 1, and a relative velocity between 0.03 m/s (the bouncing barrier estimate, in pre-chondrule forming nebular conditions \citep{Jacquet2014size}) and 1 km/s (for droplet integrity; \citet{ArakawaNakamoto2019}). The oxygen fugacity constraints are taken from \citet{Tenneretal2015}, which strictly constrain the solid/gas ratio, and thus depends on an assumed gas density, here the midplane Minimum Mass Solar Nebula (MMSN), only for reference (of course this profile does not lead to the pressure bumps advocated in the text). The MMSN midplane dust density is also plotted, as well as a dust-enriched version with dust/gas (mass) ratio of unity conducive to streaming instability, and the Roche density threshold conducive to gravitational collapse.}
\label{densities}
\end{figure}

  At the order-of-magnitude level, this is consistent (or may be brought in consistency, see \ref{Mg}) with constraints from limited Mg isotopic fractionations of chondrules, as well as oxygen fugacity and compound chondrule frequencies as depicted in Fig. \ref{densities}. Such elevated densities are needed to stabilize liquids \citep{CuzziAlexander2006}. \citet{EbelGrossman2000} quote minimum solid/gas enrichments of 12.5 and 425 at 10$^{-3}$ and 10$^{-6}$ bar; corresponding to $\rho_p \approx 10^{-6}\:$kg/m$^3$ and $4\times 10^{-8}\:$kg/m$^3$ (after condensation of all rock-forming elements), respectively.
  
  Yet, astrophysically speaking, this represents a huge value, comparable to the \textit{gas} density at 1 AU for a Minimum Mass Solar Nebula \citep{Hayashi1981}, although the solar solid/gas (mass) ratio is $10^{-2}$ (Fig. \ref{densities}). Solid/gas ratios of order unity, as suggested by fayalite contents in type I (and a fortiori type II) chondrules \citep[e.g.][]{Tenneretal2015}, would require efficient settling to the midplane, and could be conducive to streaming instabilities. 
This could be assisted by radial concentration in pressure bumps \citep[e.g.][]{Jacquetetal2024}. So nebular chondrule-forming scenarios remain viable. At any rate, chondrule-forming densities may not be reached beyond a few AU (Fig. \ref{densities}), in line with the rarity of chondrules in CI chondrites and C asteroids \citep[e.g.][]{Morinetal2022,Nakamuraetal2023,Laurettaetal2024}. 

\subsection{NC/C comparison}
  The previous subsection has addressed data from carbonaceous chondrites. However, chondrules in the NC superclan seem largely autochtonous to it, judging from their nucleosynthetic isotopic anomalies \citep[e.g.][]{Schneideretal2020}, and thus formed in a reservoir distinct from the former.  How do their alkali elemental and isotopic systematics compare?

  In terms of elemental abundances, the NC chondrules have on average higher alkali abundances than their C counterparts, more or less reflected by the bulk chondrite abundances. The difference narrows though if we restrict attention to type I chondrules (that is, similar oxygen fugacities relative to IW), as in table \ref{table alkali type I}. While EH (type I) chondrules show near chondritic alkali abundances, OC are only a factor of $\sim$1-3 above their CC counterparts. OC, CMO and CR chondrules show overlapping alkali vs. SiO$_2$ trends after averaging (Fig. \ref{Na_vs_SiO2}, \ref{K_vs_SiO2}).

\begin{table}
\caption{Averaged alkali abundances of type I chondrules. The data are CI-normalized \citep{Lodders2003} and based on the averaged composition of chondrule mesostases in our compilation (number $n$ of analyses specified)and errors are one standard error of the mean. The CR average does not include the 50 \citet{Mathieu2009} measurements which seem to have preferentially targeted a few alkali-rich objects.}
\label{table alkali type I}
\begin{tabular}{c c c c}
\hline
Group & Na/Al & K/Al & $n$\\
\hline
EH & 1.0$\pm$0.1 & 1.1$\pm$0.2  & 56\\
OC & 0.30$\pm$0.02 & 0.23$\pm$0.02  & 171\\
CMO & 0.19$\pm$0.02 & 0.19$\pm$0.02 & 75\\
CR & 0.13$\pm$0.02 & 0.26$\pm$0.09 & 58\\
CHB & 0.16$\pm$0.02 & 0.09$\pm$0.02 & 39\\
\hline
\end{tabular}
\end{table}

  In terms of isotopic composition,  OC chondrules measured by \citet{KoefoedWang2025} average $\delta^{41/39}$K=-1 \textperthousand, which, if tied to the same (CI) chondrule-forming reservoir composition, would yield an MDF three times that of carbonaceous chondrites. Since enstatite chondrites exhibit bulk potassium isotopic ratios intermediate between bulk ordinary chondrites and extrapolated CC chondrules, the MDF of their chondrules should be intermediate as well. Likewise, \citet{Wangetal2025} measured an average $\delta^{87/85}$Rb=-0.12$\pm$0.02\textperthousand$\:$ for EH3 chondrites, which, if identified with the chondrule composition, would yield an MDF double that of CC chondrules. Semarkona chondrules yield an average $\delta^{7/6}$Li=0.4\textperthousand$\:$ not much heavier than Allende chondrules \citep{Seitzetal2012}.

  Given the orders of magnitude over which $\rho_p$ might be expected to vary between the inner and the outer disk (in an MMSN, for example, it scales like $R^{-2.75}$ with $R$ the heliocentric distance; see Fig. \ref{densities}), it is surprising the elemental abundances and estimated MDF (both dependent on $\rho_p/|dT/dt|$) only differ by a factor of a few. "Planetary" scenarios where chondrule-forming conditions (e.g. an impact plume) are decoupled from the surrounding disk could find comfort in such evidence (even though that from precursor composition would remain problematic; \citet{Marrocchietal2024ISSI}). There is however some evidence that cooling was slower for the CC chondrules. Most chondrule mesostases in CV \citep{BrearleyJones1998} and CR \citep{Harjuetal2014} chondrites are holocrystalline, not glassy, and about 10 \% of chondrules in CO chondrites are plagioclase-bearing, which the \citet{WickJones2012} experiments only reproduced for their slowest cooling rate of 1 K/h. The suprachondritic Rb/K ratios reported in the alkali-poorest Allende chondrules discussed previously are not seen in ordinary chondrites, with chondritic \citep{Alexander1994} or subchondritic \citep{Grossmanetal2007LPSC} values being measured in LL3 chondrites, so this would suggest higher equilibration temperatures. To achieve similar $\rho_p/|dT/dt|$ ratios, faster cooling rates for NC chondrules would mean higher chondrule-forming densities, in line with their shorter heliocentric distances of formation. We note that $\delta^{41/39}$K (and thus MDF) does not seem to change between type I and (faster cooled; \citet{Jacquetetal2015LL}) type II chondrules in OCs \citep{KoefoedWang2025}, so this approximate compensation  between chondrule-forming densities and cooling rates also applied within single chondrule-forming reservoirs.

  Why would there be such a compensation? Perhaps the cooling rate was dictated by the exit from the hot chondrule-forming region, whose "boundary" may have had an optical depth of order unity. Since the opacity would have been proportional to $\rho_p$, for a given relative velocity, the timescale of the exit would have been inversely proportional to it, and thus the cooling rate would have been proportional to $\rho_p$. This is only speculative at this point; this compensation would remain a constraint on specific chondrule-forming processes \citep{Jacquetetal2015LL}.

\subsection{Perspectives}
  In this contribution, we have offered a synthesis on the incorporation of alkali elements in chondrules, undertaking to account for bulk composition, mesostasis and glass inclusions, olivine zoning, isotopic compositions and the alkali-zoned chondrules. However, this work is not a full modelling of the evolution of chondrules in a gas. It is rather a collection of theorems based on broad assumptions, independent of details of such a modelling (e.g. specific cooling curves), and expressed as a function of parameters (e.g. condensation coefficients, crystal fraction etc.) which are not explicitly modelled (and could vary with time), though constrained in order of magnitude. Certainly, real \textit{ab initio} kinetic calculations on the evolution of melt compositions in interaction with the gas in dust-enriched systems would be desirable (if only to test the validity of the assumptions), first in an homogeneous medium ("0D"), later perhaps introducing some spatial heterogeneities. We also encourage more cooling rate constraints on chondrules from both C and NC superclans, bearing in mind the possibility (advocated here) of nonlinear (and more specifically, accelerating) cooling. More measurements of different alkalis (e.g. the Rb/K ratio) of chondrules in different chondrite groups would be desirable. We also encourage more experiments or theoretical treatment on the condensation coefficients of alkalis and other moderately volatile elements to help model all those simultaneously beyond the oder-of-magnitude level which we have had to be content with.

\section*{Acknowledgments}
This work was supported by ANR PERSEID (ANR-25-CE49-3880). We thank Alessandro Morbidelli and Roger Hewins for discussions. The paper is dedicated to the memory of the first author's great-aunt Claire Passe and great-uncle Jean Passe.

\appendix
\section{Calculation of activity coefficients}
\label{appendix_gamma}

The activity of NaO$_{0.5}$ in the melt may be written as:

\begin{equation}
a_{\rm NaO_{0.5}}=\sqrt{a_{\rm Na_2O}}\equiv\gamma_{\rm NaO_{0.5}}x_{\rm NaO_{0.5}}
\end{equation}

with $\gamma_{\rm NaO_{0.5}}$ the activity coefficient and
\begin{equation}
x_{\rm NaO_{0.5}}
=\frac{m_l}{m_{\rm Na}}\mathrm{Na}_l
\end{equation}
the molar fraction, expressed here as a function of the mass fraction in the liquid Na$_l$ 
 and the mean "molecular" mass $m_l$. So as to fit with \citet{Mathieu2009,Mathieuetal2011}, except for the trace alkalis (as only the single-cation "molecules" satisfy Henry's law), the "molecules" in question are: NaO$_{0.5}$,  KO$_{0.5}$, MgO, Al$_2$O$_3$, SiO$_2$, P$_2$O$_5$, CaO, TiO$_2$, Cr$_2$O$_3$, MnO, FeO. So in this convention, $m_l$ is of order 60 g/mol (roughly the molecular mass of SiO$_2$).

  Judging from the Na$_2$O (mol\%) vs. $P_{\rm Na}$ diagram (Fig. V.18) of \citet{Mathieu2009}) 
, chondrule mesostasis compositions should be in the Henry domain. The applicable $\gamma_{\rm NaO_{0.5}}$ would then be calculable from the initial slopes $a_{\rm Mathieu}$ in this diagram, using the action mass law, as\footnote{In the limit of vanishing sodium contents, the molar fraction of Na$_2$O (all other above molecular species equal) is $x_{\rm NaO_{0.5}}/2$}:

\begin{equation}
\gamma_{\rm NaO_{0.5}}=\frac{f_{O_2}^{1/4}}{2K_{\rm Na}a_{\rm Mathieu}P_0}
\end{equation}

From table V.5 of \citet{Mathieu2009} for experiments at 1400 °C (and the NNO buffer at $f_{O_2}=2.11\times 10^{-6}$), we calculate the values tabulated in Table \ref{table gamma Na}. These are a function of the (volatile-free) optical basicity, which is the average (weighted by oxygen atoms) of the pure oxide values tabulated in table \ref{table lambda} below \citep{Mathieu2009}:

\begin{table}
\caption{Optical basicity}
\label{table lambda}
\begin{tabular}{c c c c c c c c c}
\hline
oxide & MgO & Al$_2$O$_3$ & SiO$_2$ & P$_2$O$_5$ & CaO & TiO$_2$ & MnO & FeO\\
$\Lambda$ & 0.78 & 0.6 & 0.48 & 0.4 & 1 & 0.61 & 1 & 1\\
\hline
\end{tabular}
\end{table}

Mesostasis values have then been estimated using a Lagrange polynomial interpolation of the logarithms of its values at $\Lambda$=0.48, 0.55 and 0.60. For (more optically basic) bulk silicate chondrules, the interpolation was for $\Lambda$=0.55, 0.60 and 0.65.

\begin{table}
\caption{Activity coefficient for NaO$_{0.5}$ at 1400 °C}
\label{table gamma Na}
\begin{tabular}{c c c c c c}
\hline
$\Lambda$ & 0.48 & 0.55 & 0.60 & 0.65 & 0.72\\
$\gamma_{\rm NaO_{0.5}}$ &  $3.8\times 10^{-5}$ & $8\times 10^{-5}$ & $1.5\times 10^{-4}$ & $2\times 10^{-3}$ & $2\times 10^{-2}$\\
\hline
\end{tabular}
\end{table}

  These are activity coefficients for a specific temperature (1673 K
).  For a regular solution, $\mathrm{ln}\gamma\propto T^{-1}$ \citep[e.g.][]{GhiorsoSack1995}. This is essentially the behavior shown in fits for $T\ll 10^4$K \citep{Zhangetal2021,Wolfetal2023}. So we will adopt:
\begin{equation}
\gamma_{\rm NaO_{0.5}}(\chi,T)=\gamma_{\rm NaO_{0.5}}(\chi, 1673 \mathrm{K})^{\frac{\rm 1673 K}{T}}
\end{equation}
with $\chi$ symbolizing the chemical composition of the melt, and the same for the other alkalis.

\section{Thermodynamics of recondensation of other alkalis}
\label{alkali thermo}

\subsection{Potassium}
The reaction constant of
\begin{equation}
\rm KO_{0.5}(l)=K (g)+\frac{1}{4}O_2 (g)
\end{equation}
 is :
\begin{equation}
K_K=8.7\times 10^5 \mathrm{exp}\left(-\frac{\rm 26207 K}{T} \right)
\end{equation}
  At large densities, some of the vapor phase K may be in the form of KCl \citep{EbelGrossman2000}.  However, this cannot be a majority above the half-condensation temperatures of the alkalis.
  Indeed, if we let
\begin{equation}
K_1=1.4\mathrm{exp}\left(\frac{\rm 1822\: K}{T}\right)
\end{equation}
the constant (fitted for T between 1200 and 3000 K) of the reaction
\begin{equation}
\rm NaCl + K = Na + KCl,
\end{equation}
we find:
\begin{equation}
\frac{n_{\rm KCl}}{n_K}=K_1\frac{n_{\rm NaCl}}{n_{\rm Na}}\lesssim K_1\left(\frac{\rm Cl}{\rm Na}\right)_g=0.13\frac{\rm \left(Cl/Na\right)_g}{\rm \left(Cl/Na\right)_{CI}}\mathrm{exp}\left(\frac{\rm 1822\: K}{T}\right)
\end{equation}
where (Cl/Na)$_{\rm g,CI}$ are the atomic Cl/Na ratio in the gas phase and CI chondrites, respectively. This is thus $\ll 1$ so long Na has not significantly recondensed. Part of the gas phase Cl would also be locked in HCl so this is a conservative upper bound. So we might not expect more than a factor of two decrease of the K/Na ratio in the monoatomic gas relative to elemental abundances.

  Assuming both Na and K are equilibrated with the gas, the ratio of the K/Na ratios in the liquid and in the monoatomic gas is:
\begin{equation}
(\mathrm{K/Na})_{\rm chd/g}=\frac{\gamma_{\rm NaO_{0.5}}}{\gamma_{\rm KO_{0.5}}}\frac{K_{\rm Na}}{K_K}=1.1\mathrm{exp}\left(\left(1673\mathrm{ln}\frac{\gamma_{\rm NaO_{0.5}}(1673 K)}{\gamma_{\rm KO_{0.5}}(1673 K)}-4479\right) \mathrm{K}/T\right)
\end{equation}
We have no activity coefficient model for KO$_{0.5}$, as the \citet{Georges2000} parameterization of the solubility of K is only applicable for much more Ca-rich (30-45 wt\% CaO) and Al-poor compositions than chondrule mesostases, and the MELTS code yields spurious results \citep{Zhangetal2021}. For a basaltic composition, the values reported by \citet{Zhangetal2021} yield  $\gamma_{\rm NaO_{0.5}}/\gamma_{\rm KO_{0.5}}=24$  (for identical evaporation coefficients) so (K/Na)$_{\rm chd/g}=1.8$ at 1673 K
. 

\begin{figure}
\centering
\includegraphics[width=\textwidth]{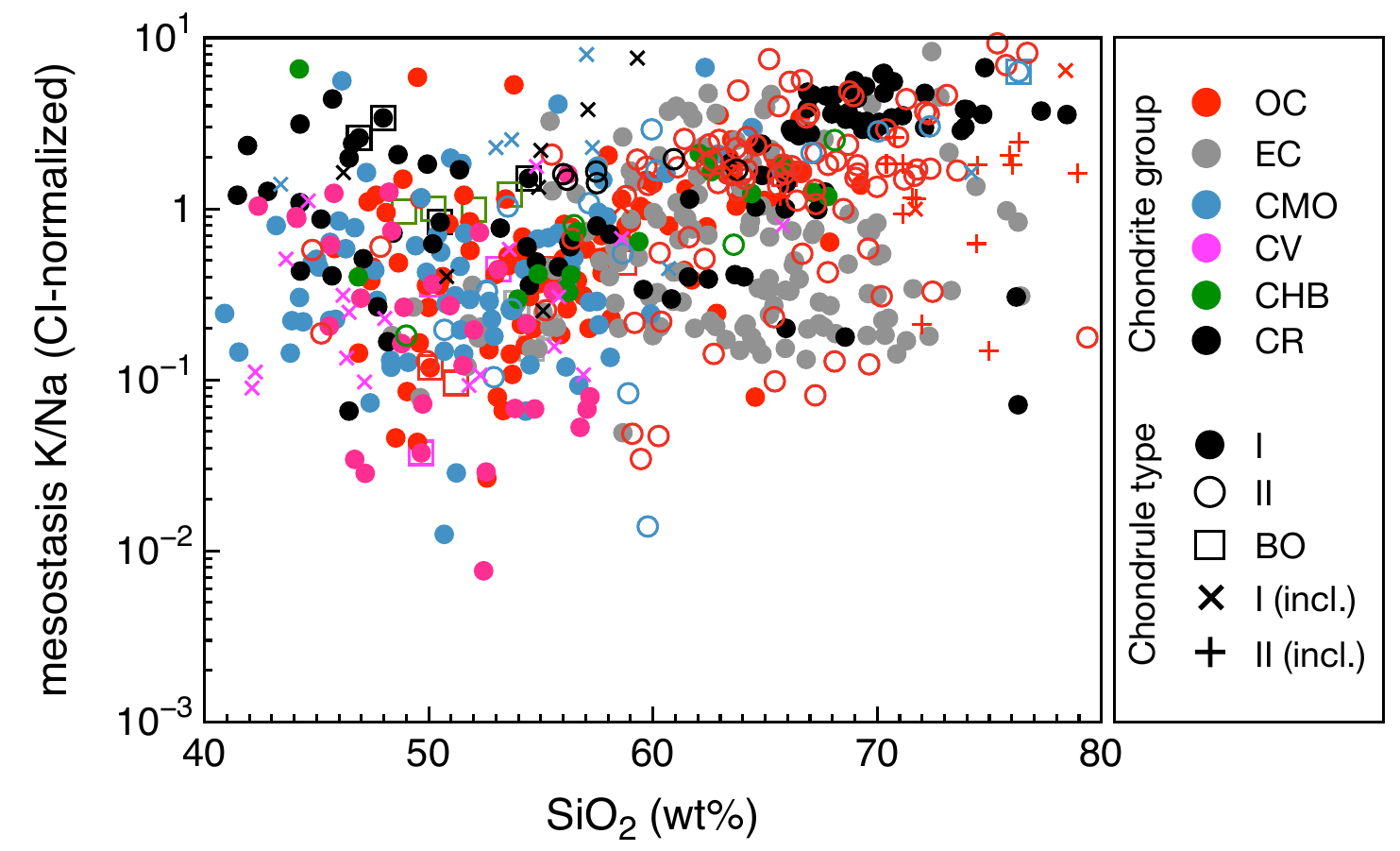}
\caption{Mesostasis K/Na ratio vs. silica.}
\label{KNa_vs_SiO2}
\end{figure}

  However, mesostasis data suggest that the most basic melts show subchondritic K/Na ratios (Fig. \ref{KNa_vs_SiO2}). This includes olivine-hosted glass inclusions and barred olivine (BO) chondrule mesostases which were presumably instantly sealed off further interaction with the gas (by shell formation upon nucleation for BO chondrules). 
This may come about because half of gaseous K was locked in KCl as mentioned above, or uncertainties in the thermodynamic data. More silica-rich melts may have higher $\gamma_{\rm NaO_{0.5}}/\gamma_{\rm KO_{0.5}}$ hence higher K/Na ratios, which could also come about by the increase of the K/Na ratio of the residual gas in K if (K/Na)$_{\rm chd/g}$ was actually below unity.

\subsection{Rubidium}
    We may likewise express the reaction constant \citep{Sossietal2019MOVE}
\begin{equation}
K_{\rm Rb}=2.1\times 10^{13}\exp{\left(-\frac{\rm 48400\:K}{T}\right)}
\end{equation}
 Thus:
\begin{equation}
(\mathrm{Rb/K})_{\rm chd/g}=\frac{\gamma_{\rm KO_{0.5}}}{\gamma_{\rm RbO_{0.5}}}\frac{K_{\rm K}}{K_{\rm Rb}}=
\exp{\left(22193\mathrm{K}\left(\frac{1}{T}-\frac{1}{\rm 1300 \: K}\right)\right)}\frac{\gamma_{\rm KO_{0.5}}}{\gamma_{\rm RbO_{0.5}}}
\end{equation}
The \citet{Zhangetal2021} experiments suggest that $\gamma_{\rm KO_{0.5}}=\gamma_{\rm RbO_{0.5}}$. So (Rb/K)$_{\rm chd/g}$ increases with decreasing temperature and becomes unity at 1300 K. 

\subsection{Lithium}
Likewise, from \citet{Sossietal2019MOVE}:
\begin{equation}
K_{\rm Li}=1.8\times 10^{7}\exp{\left(-\frac{\rm 49673\:K}{T}\right)}
\end{equation}
Thus
\begin{equation}
(\mathrm{Li/Rb})_{\rm chd/g}=\frac{\gamma_{\rm KO_{0.5}}}{\gamma_{\rm LiO_{0.5}}}\frac{K_{\rm Rb}}{K_{\rm Li}}=2\times 10^{6}\frac{\gamma_{\rm RbO_{0.5}}}{\gamma_{\rm LiO_{0.5}}}
\exp{\left(1273\mathrm{K}\left(\frac{1}{T}-\frac{1}{\rm 1673 \: K}\right)\right)}
\end{equation}
\citet{Sossietal2019MOVE} report $\gamma_{\rm LiO_{0.5}}=1.8\pm 0.2$ and $\gamma_{\rm RbO_{0.5}}=(1.3\pm 0.15)\times 10^{-4}$ in a ferrobasaltic melt at IW-8 and 1400 $^\circ$C, so this should be 180 then. Li is thus more refractory than the other alkalis. 
 
Table \ref{table TE Trec} summarises the Arrhenian temperature dependences of the alkalis.

\begin{table}
\caption{Arrhenian temperature dependences. 
The $T_{\rm rec}$ are calculated for the indicated, typical values of $\gamma_{\rm EO_{0.5}}(1673\:\mathrm{K})$.}
\label{table TE Trec}
\begin{tabular}{c c c c}
\hline
E & $T_E$ (K) & $\gamma_{\rm EO_{0.5}}(1673\:\mathrm{K})$ & $T_{\rm rec, E}$ (K)\\
\hline
Li & 49673 & 1 & 33108\\
Na & 30686 & $10^{-4}$ & 29619\\
K & 26207 & $10^{-5}$ & 28903\\
Rb & 48400 & $10^{-5}$ & 51096\\
\hline
\end{tabular}
\end{table}

\section{Mass-dependent fractionation in a closed chondrule-forming volume}
\label{appendix MDF}

  We consider a homogeneous volume containing gas and a population of identical chondrules, with N$_i$ the total number density of isotopes $i$. Then, expressing $n_i=N_i-n_pX_i$, with $n_p$ the number density of chondrules, the Hertz-Knudsen equation \ref{H-K} may be rewritten as:
\begin{equation}
\label{H-K conv}
\frac{\mathrm{d}X_i}{\mathrm{d}t}=\frac{X_{\rm conv,i}-X_{\rm i}}{t_{\rm conv, i}}
\end{equation}
where
\begin{equation}
t_{\rm conv,i}=\left(\frac{1}{t_{\rm rec,i}}+\frac{1}{\tau_{\rm eff,i}}\right)^{-1}
\end{equation}
\begin{equation}
X_{\rm conv,i}=\frac{N_i/n_p}{1+\tau_{\rm eff,i}/t_{\rm rec_i}}
\end{equation}

  The formal solution of equation \ref{H-K conv}, evaluated at the end of the cooling ($t=t_{\rm max}$) is:
\begin{eqnarray}
X_i &=&\int_0^{t_{\rm max}}\frac{\mathrm{d}e^{-x_i}}{\mathrm{d}t}X_{\rm conv,i}\mathrm{d}t+X_i(0)\mathrm{exp}\left(-\int_0^{t_{\rm max}}\frac{\mathrm{d}t}{t_{\rm conv,i}}\right)\nonumber\\
&\approx& \int_0^{t_{\rm max}} \frac{\mathrm{d}e^{-x_i}}{\mathrm{d}t}X_{\rm conv,i}\mathrm{d}t
\end{eqnarray}
with the "memory variable"
\begin{equation}
x_i\equiv\int_t^{t_{\rm max}}\frac{\mathrm{d}t'}{t_{\rm conv,i}}
\end{equation}
and where the approximation assumes the limit of forgotten beginnings (i.e. $x_j(0)\gg 1$). Of course, the same equations obtain for isotope $j$. We now assume no equilibrium isotopic fractionation, that is $X_{\rm conv,i}/X_{\rm conv,j}=N_i/N_j$ (as reported for K by \citet{Nieetal2021}). Since $x_i=x_j\sqrt{m_j/m_i}$, we have:
\begin{equation}
\frac{X_i/X_j}{N_i/N_j}-1\approx\left(1-\sqrt{\frac{m_j}{m_i}}\right)\frac{1}{X_j}\int_0^{t_{\rm max}}\frac{\mathrm{d}}{\mathrm{d}t}\left(x_je^{-x_j}\right)X_{\rm conv,j}\mathrm{d}t
\end{equation}
Thus:
\begin{equation}
\mathrm{MDF}\approx\frac{1}{2}\left(-1+\frac{1}{X_j}\int_0^{t_{\rm max}}x_je^{-x_j}\frac{X_{\rm conv,j}}{t_{\rm conv,j}}\mathrm{d}t\right)
\end{equation}
We now assume that both $X_{\rm max,j}\equiv N_j/n_p$ and $\tau_{\rm eff,j}=\left(n_p\pi a^2\alpha_j v_{T,j}(1-\phi)\right)^{-1}$ were constant, which, given that $a^2T^{1/2}\alpha_j(1-\phi)$ likely only weakly varied during recondensation, essentially amounts to constant $n_p$ and $N_j$ (i.e. no volume change of the gas+chondrules parcel). Since $X_{\rm conv,j}/t_{\rm conv,j}
=X_{\rm max,j}/\tau_{\rm eff,j}$ would have been likewise constant, we have:
\begin{equation}
X_j=X_{\rm max,j}\int_0^{t_{\rm max}/\tau_{\rm eff,j}}e^{-x_j}\mathrm{d}z\approx X_{\rm max,j}\int_0^{\infty}e^{-x_j}\mathrm{d}z
\end{equation}
where we have introduced the normalized look-back time $z=(t_{\rm max}-t)/\tau_{\rm eff,j}$ and pushed the limit of the integral over that variable to infinity (for any integrable monotonic extrapolation of the integrand before time zero) owing to the assumed limit of forgotten beginnings (effectively removing $t_{\rm max}/\tau_{\rm eff,j}$ as a parameter). Likewise:
\begin{equation}
\mathrm{MDF}=\frac{1}{2}\left(-1+\int_0^{\infty}x_je^{-x_j}\mathrm{d}z\Big/\int_0^{\infty}e^{-x_j}\mathrm{d}z\right)
\end{equation}
with
\begin{equation}
x_j=z+\int_t^{t_{\rm max}}\frac{\mathrm{d}t}{t_{\rm rec,j}}
\end{equation}
We will neglect non-Arrhenian dependences in $t_{\rm rec,j}\propto e^{T_{\rm rec}/T}$. To completely express $x_j$ and entirely determine the calculations above, we need to prescribe the time-temperature ($t$-$T$) histories. We will consider two endmember possibilities. The first is the \textit{homographic} cooling
\begin{equation}
T=\frac{T_0}{1+t/t_0}
\end{equation}  
exemplifying a slowing cooling (for $t<t_{\rm max}$, zero afterward), and the second is the \textit{inverse-log} cooling
\begin{equation}
T=-\frac{T_0}{\mathrm{ln}\left(1-t/t_{\rm max}\right)}
\end{equation}
with an accelerating cooling for $(1-e^{-2})t_{\rm max}\leq t <t_{\rm max}$\footnote{The inflection before then will not matter for we will assume that $T_0/2=T((1-e^{-2})t_{\rm max})$ is much higher than the final equilibration temperature so no memory is retained from this part.}. These functional forms are only motivated by mathematical simplicity, for Arrhenian temperature dependences translate into easily integrable exponential and power law time dependences for the homographic and inverse-log forms, respectively. 

  Indeed, for the homographic cooling, we obtain:
\begin{equation}
x_j=z+E_{\rm g/chd}(t_{\rm max})\left(e^{z/z_0}-1\right)z_0
\end{equation}
with $E_{\rm g/chd}(t_{\rm max})=\tau_{\rm eff,j}/t_{\rm rec,j}(t_{\rm max})$ the final equilibrium gas/chondrule budget ratio and
\begin{equation}
   z_0=\frac{T_0t_0}{T_{\rm rec}\tau_{\rm eff,j}}
\end{equation}

  For the inverse-log cooling, we have:
\begin{equation}
x_j=z\left(1+\frac{1}{1+T_{\rm rec}/T_0}\left(\frac{z}{z_{50}}\right)^{T_{\rm rec}/T_0}\right)
\end{equation}
with $z_{50}$ the normalized look-back time corresponding to (equilibrium) half-condensation.

  Fig. \ref{MDF_vs_C} plots the MDF for different values of $E_{\rm g/chd}(t_{\rm max})$  (homographic) and $T_{\rm rec}/T_0$ (inverse-log). The MDF are generally close to -0.5 especially for low condensed fractions. The homographic cooling only achieves low absolute MDF when
 the condensed fraction is near the equilibrium one at $t_{\rm max}$, i.e. $(1+E_{\rm g/chd}(t_{\rm max}))^{-1}$ and thus effectively imposed by a quench. The inverse-log cooling would take $T_{\rm rec}/T_0<1$ to remotely compete, which would also amount to a quench. Indeed, with $T_{\rm rec}$ being a few $10^4 K$ (\ref{alkali thermo}), the half-condensation temperature $T_{50}$ would be reached in the steep decrease part of the cooling as $T_{50}\ll T_{\rm rec}<T_0$.

\section{Effect of diffusion}
\label{diffusion}

The derivations in section \ref{kinetics} assumes that the chondrule melt is well-mixed. This appendix examines this assumption from the point of view of diffusion in the melt. (One could also envision the possibility of turbulent motions in that melt enhancing transport without any isotopic selectivity).

\subsection{Diffusion timescale}
 Generally speaking, diffusion coefficients in a melt $D$ follow an Arrhenian dependence of the form:
\begin{equation}
D=D_0 \exp{(-T_D/T)}=D(\mathrm{1673 K})\exp{\left(T_D\left(\frac{1}{\rm 1673 \: K}-\frac{1}{T}\right)\right)}
\end{equation}
It is somewhat uncertain what the relevant parameterization for chondrule mesostasis is, as $D_0$ and $T_D$ (that is, the activation energy divided by the gas constant) depend on melt composition. Yet a "compensation rule" seems to maintain D(1673 K) around $10^{-9}\:$m$^2/$s for Na \citep{Lowryetal1982,Mungall2002,Zhangetal2010} so most of the uncertainty resides on T$_D$. For other alkalis, \citet{Zhangetal2021} report similar 1673 K diffusivities of 1.05$\times 10^{-9}$ and 1.2$\times 10^{-9}\:$m$^2$/s for Rb and K in a basaltic melt. 
However, compilations by \citet{Zhangetal2010,Mungall2002} indicate lower diffusion coefficients (by 1-2 orders of magnitude) for more silica-rich compositions for those alkalis. In either case $T_D$ takes typical values of 1-2$\times 10^4$K \citep[e.g.][]{Zhangetal2010,Lowryetal1982}, with basaltic melts on the higher end of the range. As will matter later, these are lower than calculated $T_{\rm rec}$ values (table \ref{table TE Trec}; \citet{OzawaNagahara2001}).

   Since a chondrule is only partially molten, the applicable effective diffusion coefficient is reduced. Neglecting diffusion through the crystals, an empirical relationship may be given by \citep{Crank1975}:  
\begin{equation}
D_{\rm eff}=\frac{1-\phi}{1+\phi/w}D
\end{equation}
with $\phi$ the volume fraction of crystals and $w$ a dimensionless, shape-dependent parameter (2 for a sphere).  

  We define the diffusion timescale as:
\begin{eqnarray}
t_{\rm diff} &=& \frac{a^2}{\pi^2 D_{\rm eff}}\\
&=& 9\:\mathrm{s}\left(\frac{a}{\rm 0.3\: mm}\right)^2\left(\frac{1-\phi}{1+\phi/w}\right) \left(\frac{\rm 10^{-9} m^2/s}{D(1673 \mathrm{K})}\right)\exp{\left(T_D\left(\frac{1}{T}-\frac{1}{\rm 1673 \: K}\right)\right)}\nonumber
\end{eqnarray}
where the $\pi^2$ factor naturally arises in the calculation of the next subsection. (From Fig. 6.1 of \citet{Crank1975}, for a constant surface concentration with zero initial inner concentration, this is a good estimate for the time for the center to reach half that surface concentration). Since $t_{\rm rec}$ increases more steeply with cooling (as $T_{\rm rec}>T_D$), the diffusion timescale would be indeed expected to be shorter than it when alkalis condense quantitatively. This was also observed in the evaporation experiments of \citet{Tsuchiyamaetal1981}, whose chondrule analogs had uniform (more or less depleted) Na concentrations. Let us now examine more quantitatively the effect of diffusion.

\subsection{Diffusion in a sphere under recondensation}

  Assuming a partially molten chondrule devoid of alkalis at time $t=0$, the concentration $C_i(r,t)$ (atoms per unit volume) of an isotope $i$ of the melt/crystal mix may be written in terms of its surface value $C_i(a,t)$, as a function of distance to center $r<a$ and time $t$, as:
\begin{equation}
C_i(r,t)=2\sum_{n=1}^{\infty}(-1)^{n-1}\mathrm{sinc}\left(\frac{n\pi r}{a}\right)\int_0^{t}\frac{\partial e^{-n^2 x_{D,i}}}{\partial t'}
C_i(a,t')\mathrm{d}t'
\end{equation}  
where the equation 6.23 of \citet{Crank1975} has been adapted to a time-varying (yet spatially uniform) $D_{\rm eff,i}$ with the diffusion-related memory variable
\begin{equation}
x_{D,i}\equiv
\int_{t'}^t\frac{\mathrm{d}t''}{t_{\rm diff,i}}
\end{equation}
Thus, neglecting any radius change upon recondensation of the alkalis, its volume-weighted average is:
\begin{eqnarray}
\label{Ci avg}
\overline{C}_i &=&\frac{6}{\pi^2}\sum_{n=1}^{\infty}\frac{1}{n^2}\int_0^{t}\frac{\partial e^{-n^2 x_{D,i}}}{\partial t'}
C_i(a,t')\mathrm{d}t'\nonumber\\
&=& C_i(a,t)-\frac{6}{\pi^2}\sum_{n=1}^{\infty}\frac{1}{n^2}\int_0^{t}e^{-n^2 x_{D,i}}
\frac{\mathrm{d}C_i(a,t')}{\mathrm{d}t'}\mathrm{d}t'\nonumber\\
&\approx & C_i(a,t)-\frac{\pi^2}{15}t_{\rm diff,i}\frac{\mathrm{d}C_i(a,t)}{\mathrm{d}t}
\end{eqnarray}
where we neglect $C_i(a,0)$ (the limit of forgotten beginnings) and the final equality assumes
$\frac{d}{dt}\left(t_{\rm diff,i}\frac{d}{dt}C_i(a,t)\right)/\frac{d}{dt}C_i(a,t)\ll 1$. 
Thus the Hertz-Knudsen equation
\begin{equation}
\label{H-K surface}
\frac{\mathrm{d}\overline{C}_i}{\mathrm{d}t}=\frac{C_{\rm buf,i}-C_i(a,t)}{t_{\rm rec,i}}
\end{equation}
with $C_{\rm buf,i}\equiv 3X_{\rm buf,i}/(4\pi a^3)$, is a closed integro-differential equation in $C_i(a,t)$.

\subsection{Diffusion- vs. recondensation-limited near-equilibrium}
\label{diffusion-limited near eq}

From equations \ref{Ci avg} and \ref{H-K surface}, in the limit of short $t_{\rm diff, rec}$, we have:
\begin{equation}
\overline{C}_i\approx C_{\rm buf,i}-\left(t_{\rm rec,i}+\frac{\pi^2}{15}t_{\rm diff,i}\right)\frac{\mathrm{d}C_{\rm buf,i}}{\mathrm{d}t}
\end{equation}
Thus, if $D_{\rm eff,i}\equiv D_{\rm eff,j}\left(m_j/m_i\right)^\zeta$, we have:
\begin{equation}
\frac{X_i}{X_j}=\frac{\overline{C}_i}{\overline{C}_j}\approx\frac{X_{\rm buf,i}}{X_{\rm buf,j}}\left(1-\left(t_{\rm rec,j}\left(\sqrt{\frac{m_i}{m_j}}-1\right)+\frac{\pi^2}{15}t_{\rm diff,j}\left(\left(\frac{m_i}{m_j}\right)^\zeta-1\right)\right)\frac{\mathrm{dln}X_{\rm buf,j}}{\mathrm{d}t}\right)
\end{equation}
Neglecting any equilibrium isotopic fractionation, and assuming a reservoir of identical composition chondrules, we thus calculate:
\begin{equation}
\mathrm{MDF}\approx-\left(1-f_j\right)\left(\frac{f_j}{2}\tau_{\rm eff,j}+(1-f_j)\frac{\pi^2}{15}\zeta t_{\rm diff,j}\right)\frac{\mathrm{dln}E_{\rm chd/g}}{\mathrm{d}t}
\end{equation}
\citet{Zhangetal2021} report $\zeta$ values of 0.07 and 0.04 for K and Rb (although for Li it may reach 0.2 \citep{Lesher2010}). 
So the effect of finite $t_{\rm diff}$ is only isotopically appreciable if at least one order of magnitude above $t_{\rm rec}$. Even in that regime, -MDF will provide an upper bound on the value of $0.5f_j(1-f_j)\tau_{\rm eff,j}\mathrm{dln}E_{\rm chd/g}/\mathrm{d}t=0.5(1-f_j)^2 t_{\rm rec,j}\mathrm{dln}E_{\rm chd/g}/\mathrm{d}t$ discussed in section \ref{CFR densities}. 

  So how important was diffusion for chondrules? If we write\footnote{This neglects non Arrhenian dependences, as in \ref{appendix MDF}, or rather, makes the "t$_{\rm rec}$(1673 K)" an Arrhenian extrapolation with non-Arrhenian prefactors evaluated at the real temperature.} 
\begin{equation}
t_{\rm rec}=t_{\rm rec}(1673\:\mathrm{K})\mathrm{exp}\left(T_{\rm rec}\left(\frac{1}{T}-\frac{1}{1673\:\mathrm{K}}\right)\right),
\end{equation}
we can eliminate the Arrhenian factor in the diffusion/recondensation timescale ratio as follows:
\begin{eqnarray}
\frac{t_{\rm diff,j}}{t_{\rm rec,j}}&=&\frac{t_{\rm diff,j}(1673\:\mathrm{K})}{t_{\rm rec,j}(1673\:\mathrm{K})}\mathrm{exp}\left(\left(T_D-T_{\rm rec,j}\right)\left(\frac{1}{T}-\frac{1}{\rm 1673 K}\right)\right)\\
&=& \frac{t_{\rm diff,j}(1673\:\mathrm{K})}{t_{\rm rec,j}(1673\:\mathrm{K})}\left(\frac{t_{\rm rec,j}(1673\:\mathrm{K})}{t_{\rm rec,j}}\right)^{1-T_D/T_{\rm rec,j}}
\end{eqnarray}
(The exponent is positive since $T_D$ is smaller than $T_{\rm rec}$ as mentioned in the first subsection). If indeed negligible in the MDF, we have:
\begin{eqnarray}
\label{tdiff/trec MDF}
\frac{t_{\rm diff,j}}{t_{\rm rec,j}}&=& t_{\rm diff,j}(1673\:\mathrm{K})t_{\rm rec,j}(1673\:\mathrm{K})^{-T_D/T_{\rm rec,j}}\left(\frac{\left(1-f_j\right)^2}{\rm 2|MDF|}\frac{\mathrm{dln}E_{\rm chd/g}}{\mathrm{d}t}\right)^{1-T_D/T_{\rm rec,j}}\\
&=& \Big[10^{-3} 
\left(\frac{t_{\rm diff,j}(1673\:\mathrm{K})[t_{\rm diff,j}(1673\:\mathrm{K})/t_{\rm rec,j}(1673\:\mathrm{K})]^{1/(T_{\rm rec,j}/T_D-1)}}{\rm 10\: s}\right)
\nonumber\\
&&\left(\frac{0.01}{\rm |MDF|}\right)\left(1-f_j\right)^2   \left(\frac{\rm 100 K}{|\mathrm{d}T/\mathrm{dlnE_{chd/g}}|}\right)\left(\frac{|dT/dt|}{\rm 10 \:K/h}\right)\Big]^{1-T_D/T_{\rm rec,j}}\nonumber 
\end{eqnarray}
Diffusion is negligible for isotopic effects if $\zeta t_{\rm diff,j}/t_{\rm rec,j}\ll 1$. This holds for elements diffusing as fast as sodium, or within one or two orders of magnitude thereof, as seems to be the case for K and Rb \citep{Zhangetal2021} as mentioned previously. 

\subsection{Diffusion under perfect surface-gas equilibrium}
\label{Teff diff}

  The above assumes near-equilibrium. Yet we need to know whether a small |MDF| necessarily means a quench (in near-equilibrium condition) if diffusion is the rate-limiting step, instead of recondensation examined in \ref{appendix MDF}. We now assume perfect surface equilibration with the gas, i.e. that the surface concentration is buffered at $C_{\rm buf}$. As in \ref{appendix MDF}, we consider a purely Arrhenian dependence of $E_{\rm chd/g}=t_{\rm rec,j}/\tau_{\rm eff,j}\propto e^{T_{\rm rec}/T}$ such that
:
\begin{equation}
C_{\rm buf,i}=\frac{C_{\rm max,i}}{1+\mathrm{exp}\left(T_{\rm rec}\left(\frac{1}{T_{50}}-\frac{1}{T}\right)\right)}
\end{equation}
 where we have introduced the 50 \% condensation temperature $T_{50}$. Unlike \ref{appendix MDF}, this treats the ambient gas partial pressures as given, without thus (explicitly) modeling the feedback of incomplete equilibration of the chondrule population on it. Yet the error on isotopic effects will be limited for the small recondensation fractions of interest for alkalis. 

  Equation \ref{Ci avg} 
 may then be rewritten as:
\begin{equation}
\overline{C}_i=\frac{6}{\pi^2}\int_0^{\infty}\sum_{n=1}^{\infty}e^{-n^2 x_{D,i}}C_{\rm buf,i}(x_{D,i})\mathrm{d}x_{D,i}
\end{equation}
 As in \ref{appendix MDF}, we will consider the homographic and inverse-log $T-t$ cooling laws. Thus:
\begin{eqnarray}
x_{D,i}=
\int_t^{+\infty}\frac{\mathrm{d}t}{t_{\rm diff,i}}
=\left(\frac{\pi}{a}\right)^2\frac{\mathrm{d}t}{\mathrm{d}D_{\rm eff,i}^{-1}}\times
\left\{
\begin{array}{ll}
1 & \mathrm{(homographic)}\\
\left(1+T_0/T_D\right)^{-1} &\mathrm{(inverse-log)}
\end{array}
\right. 
\end{eqnarray}
 
In both cases, we may express:
\begin{equation}
C_{\rm buf,i}=\frac{C_{\rm max,i}}{1+\left(x_{D,i}/x_{50,i}\right)^\beta}
\end{equation}
with $x_{50,i}$ the value of $x_{D,i}$ at $T=T_{50}$ and
\begin{eqnarray}
\beta=
\left\{
\begin{array}{ll}
T_{\rm rec}/T_D & \mathrm{(homographic)}\\
 T_{\rm rec}/\left(T_D+T_0\right) &\mathrm{(inverse-log)}
\end{array}
\right. 
\end{eqnarray}

\begin{figure}
\centering
\includegraphics[width=\textwidth]{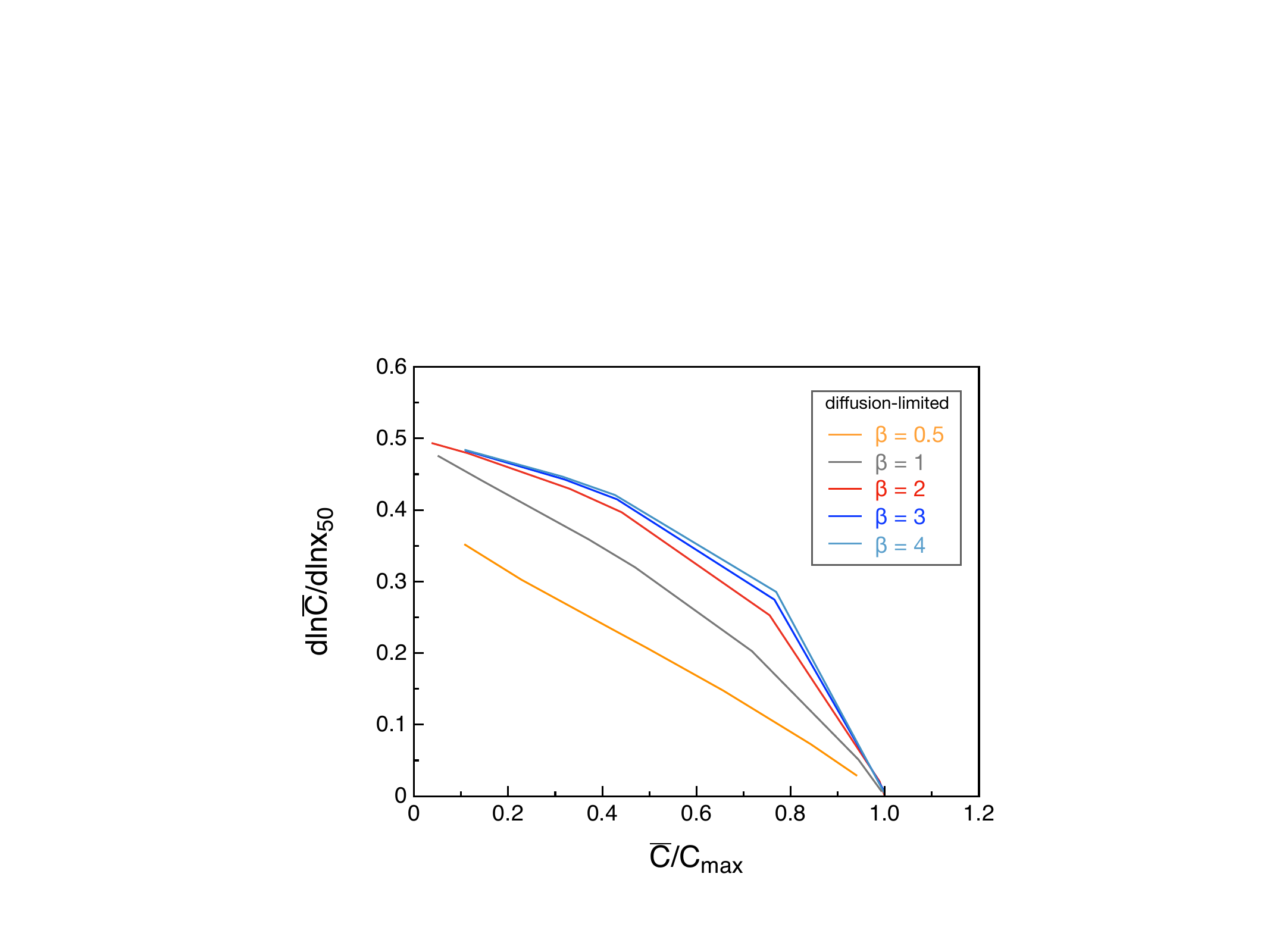}
\caption{dln$\overline{C}/$dln$x_{50}$ as a function of $\overline{C}/C_{\rm max}$.}
\label{dlnC_vs_C}
\end{figure}

  In terms of isotopic fractionation, since $x_{50,i}=(m_j/m_i)^\zeta x_{50,j}$, we have:
\begin{equation}
\frac{\overline{C}_iC_{\rm max,j}}{\overline{C}_jC_{\rm max,i}}-1\approx \frac{\mathrm{dln}\overline{C}}{\mathrm{dln}x_{50}}\left(\left(\frac{m_j}{m_i}\right)^\zeta-1\right)
\end{equation}
with 
\begin{equation}
\frac{\mathrm{dln}\overline{C}}{\mathrm{dln}x_{50}}=\frac{\beta}{x_{50}^\beta}\sum_{n=1}^{\infty}\int_0^{+\infty}\frac{x^\beta e^{-n^2 x}\mathrm{d}x}{\left(1+(x/x_{50})^\beta\right)^2} \left( \sum_{n=1}^{\infty}\int_0^{+\infty}\frac{e^{-n^2 x}\mathrm{d}x}{1+(x/x_{50})^\beta}      \right)^{-1}
\end{equation}
This vanishes for $x_{50}\gg 1$ (near-quantitative condensation) and converges toward 0.5 for $x_{50}\ll 1$ (limited recondensation)\footnote{Change the integration to variable $y=x/x_{50}$, approximate the Riemann sum $\sum_{n=1}^{\infty}e^{-n^2 x_{50}y}\sim\sqrt{\pi/(x_{50}y)}/2$ and perform an integration by part of the numerator by recognizing $d(1+y^\beta)^{-1}/dy$.}, hence a MDF of $-\zeta/2$ (Fig. \ref{dlnC_vs_C}). Below half-condensation, very low $\beta$ would be required to significantly lower |MDF|. These would correspond to high $T_0$ in an inverse-log T-t law, hence effectively a quench regime. 

  The observed |MDF| (table \ref{table MDF}) are much smaller than $\zeta/2$ for $\zeta$ values of 0.07 and 0.04 for K and Rb \citep{Zhangetal2021}. So even if diffusion were the rate-limiting step, a quench would still be required.

\section{Comparison with Mg isotopic constraints}
\label{Mg}

Allende bulk chondrule $\delta^{25}$Mg anticorrelates with Mg/Al \citep{Galyetal2000}, suggestive of Rayleigh fractionation upon evaporation, if much more limited than the vacuum prediction. \citet{CuzziAlexander2006} inferred thence $\rho_p\approx 7-13\times 10^{-6}$ kg/m$^3$ to approach Mg saturation sufficiently, consistent with our conclusions, but what they really estimated was the quantity
\begin{equation}
a_{\rm Cuzzi}\equiv\frac{t_h}{\tau_{\rm eff,Mg}}
\end{equation}
with $t_h$ the heating time
.

  Since
\begin{equation}
\mathrm{MDF}_K=-\frac{1}{2}f_K\left(1-f_K\right)\tau_{\rm eff,K}\frac{\rm dlnK_{chd/g}}{\mathrm{d}t}
\end{equation}
we have
\begin{eqnarray}
\mathrm{MDF}_K a_{\rm Cuzzi}&=&-\frac{1}{2}\frac{\alpha_{\rm Mg}}{\alpha_K}\sqrt{\frac{m_K}{m_{\rm Mg}}}f_K\left(1-f_K\right)t_h\frac{\rm dlnK_{chd/g}}{\mathrm{d}t}\left(1-\phi\right)^{-1}\nonumber\\
&=& -0.13 
\left(\frac{\alpha_{\rm Mg}}{\alpha_{\rm K}}\right)\frac{f_K(1-f_K)}{0.2}t_h\frac{\rm dlnK_{chd/g}}{\mathrm{d}t}\left(1-\phi\right)^{-1}
\end{eqnarray}
The left-hand-side evaluates to -0.03 from table \ref{table MDF} and the $a_{\rm Cuzzi}=6\pm 1$ obtained by \citet{CuzziAlexander2006}. It would require a $t_h$ of order the cooling through one thermal e-fold of K condensation (about 100 K). It seems then difficult to ignore isotopic exchange in the cooling stage. To recover the isotopic fractionations observed, we would need lower densities than calculated by \citet{CuzziAlexander2006} to achieve greater isotopic fractionation before cooling damped them.

  The isotopically heavy Mg of the high Al/Mg chondrules of \citet{Galyetal2000} are however not necessarily related to evaporation \textit{during} chondrule formation. In fact, their Al/Mg ratio may not be either. Indeed, \citet{Gerberetal2017} found that refractory elements of CV and CR chondrules correlated with their Ti isotopic anomalies, approaching those of CAIs in proportion, and Al-rich chondrules often show REE patterns reminiscent of diluted CAIs \citep{MisawaNakamura1988,JacquetMarrocchi2017,Zhangetal2020}. If the two most Al-rich chondrules (ARC) of \citet{Galyetal2000}, now at an average $\delta^{25}$Mg=2.55\textperthousand, inherited a $\delta^{25}$Mg around 10 \textperthousand$\:$ from CAI-rich precursors and then exchanged with the gas to approach a buffer value around the Allende bulk value of 1.59 \textperthousand$\:$ with an e-folding timescale $\sim t_{\rm rec, Mg}$ (see equation \ref{d ratio/dt}), this would have required an exchange time $t_{\rm exch, Mg}$ 
 given by:
\begin{equation}
a_{\rm new}\equiv \frac{t_{\rm exch, Mg}}{t_{\rm rec, Mg}}= \mathrm{ln}\left(\frac{10-1.59}{2.55-1.59}\right)=2 
\end{equation}
This is only an order-of-magnitude estimate, ignoring variations of $t_{\rm rec, Mg}$ (which otherwise would have to be replaced by its harmonic time average). And thus:
\begin{eqnarray}
\mathrm{MDF}_K a_{\rm new}&=&-\frac{f_{\rm evap, Mg}}{2} \frac{\rm \overline{Mg}_{max}}{\rm Mg_{ARC}}\frac{\alpha_{\rm Mg}}{\alpha_K}\sqrt{\frac{m_K}{m_{\rm Mg}}}f_K\left(1-f_K\right) t_{\rm exch, Mg}\frac{\rm dlnK_{chd/g}}{\mathrm{d}t}\left(1-\phi\right)^{-1}\\
&=& -0.13 f_{\rm evap, Mg}\left(\frac{\rm \overline{Mg}_{max}}{\rm Mg_{ARC}}\right)\left(\frac{\alpha_{\rm Mg}}{\alpha_{\rm K}}\right)\frac{f_K(1-f_K)}{0.2}t_{\rm exch, Mg}\frac{\rm dlnK_{chd/g}}{\mathrm{d}t}\left(1-\phi\right)^{-1}\nonumber
\end{eqnarray}
with Mg$_{ARC}$ the buffered concentration of Mg in the ARC, $\overline{\rm Mg}_{\rm max}$ the maximum average Mg concentration of the chondrules in the reservoir (that is, if Mg were all condensed) and $f_{\rm evap, Mg}$ the evaporated fraction of Mg. The left-hand-side evaluates to -0.01, but may be brought in consistency with the right-hand-side by a low $f_{\rm evap, Mg}$. Indeed, Mg is more refractory than K, and the low fraction of Mg in the gas phase allows preservation of (diluted) precursor isotopic signatures, while chondrule K has "forgotten" its beginnings.


\bibliographystyle{elsarticle-harv} 
\bibliography{alkali_els}







\end{document}